\documentclass[chicago]{emulateapj}
\usepackage{amsmath}
\usepackage{graphicx}

\slugcomment{ApJ Accepted; 2013 April 26}
\shortauthors{}

\begin{document}

\title{ISM Processing in the Inner 20 pc in Galactic center}

\author{Hauyu Baobab Liu\altaffilmark{1}}  
\author{Paul T. P. Ho\altaffilmark{1,2}} 
\author{Melvyn C. H. Wright\altaffilmark{3}} 
\author{Yu-Nung Su \altaffilmark{1}} 
\author{Pei-Ying Hsieh \altaffilmark{1,4}} 
\author{Ai-Lei Sun \altaffilmark{5}} 
\author{Sungsoo S. Kim \altaffilmark{6}} 
\author{Young Chol Minh \altaffilmark{7}}

\affil{$^{1}$Academia Sinica Institute of Astronomy and Astrophysics, P.O. Box 23-141, Taipei, 106 Taiwan}\email{hyliu@asiaa.sinica.edu.tw}

\affil{$^{2}$Harvard-Smithsonian Center for Astrophysics, 60 Garden Street, Cambridge, MA 02138}

\affil{$^{3}$Radio Astronomy Laboratory, University of California, Berkeley 601 Campbell Hall, Berkeley, CA 94720, USA}

\affil{$^{4}$Graduate Institute of Astronomy, National Central University, No. 300, \\Jhongda Rd, Jhongli City, Taoyuan County 32001, Taiwan (R.O.C.)}

\affil{$^{5}$Department of Astrophysical Sciences, Peyton Hall, Princeton University, Princeton, NJ, 08544, USA}

\affil{$^{6}$Dept. of Astronomy and Space Science, Kyung Hee University, Yongin-shi, Kyungki-do 446-701, Korea}

\affil{$^{7}$Korea Astronomy and Space Science Institute (KASI), 776 Daeduk-daero, Yuseong, Daejeon 305-348, Korea}


\begin{abstract}
We present the Submillimeter Array (SMA) 157-pointing mosaic in 0.86 mm dust continuum emission with 5$''$.1$\times$4$''$.2 angular resolution, and the NRAO Green Bank 100m Telescope (GBT) observations of the CS/C$^{34}$S/$^{13}$CS 1-0 and SiO 1-0 emission with $\le$20$''$$\times$18$''$ angular resolution. 
The dust continuum image marginally resolves at least several tens of 10-10$^{2}$ $M_{\odot}$ dense clumps in the 5$'$ field including the circumnuclear disk (CND) and the exterior gas streamers. There is very good agreement between the high resolution dust continuum map of the CND and all previous molecular line observations.  As the dust emission is the most reliable optically thin tracer of the mass, free from most chemical and excitation effects, we demonstrate the reality of the abundant localized structures within the CND, and their connection to external gas structures.  From the spectral line data, the velocity dispersions of the dense clumps and their parent molecular clouds are $\sim$10-20 times higher than their virial velocity dispersions. This supports the idea that the CND and its immediate environment may not be stationary or stable structures.  
Some of the dense gas clumps are associated with 22 GHz water masers and 36.2 GHz and 44.1 GHz CH$_{3}$OH masers. However, we do not find clumps which are bound by the gravity of the enclosed molecular gas.
Hence, the CH$_{3}$OH or H$_{2}$O maser emission may be due to strong (proto)stellar feedback, which may be dispersing some of the gas clumps.  
\end{abstract}



\clearpage
\section{Introduction }
\label{chap_introduction} 
The Galactic center (see Morris \& Serabyn 1996 and Mezger et al. 1996 for reviews) is a fertile ground for studying the interplay between the supermassive black hole (SMBH; see Genzel et al. 1997 and Ghez et al. 2005) and the surrounding interstellar medium (ISM).
The overall dynamics, the clumpiness, and the local kinematics may provide clues as to how the SMBH is fed by the ISM, how the circumnuclear gas streams evolve, and how the molecular cores and the young massive-stellar objects (Krabbe et al. 1991; Launhardt et al. 2002; Pfuhl et al. 2011, Lu et al. 2013, Do et al. 2013, etc) form and migrate in the Galactic center.
Previous molecular line observations indicated that the warm and turbulent gas clouds or streamers in the central $\sim$20 pc in Galactic center (Ho et al. 1991; Okumura 1991; Coil \& Ho 1999, 2000; McGary et al. 2001; Herrnstein \& Ho 2002, 2005; Oka et al. 2011) are well connected with the 2-4 pc circumnuclear disk (CND; G\"{u}sten et al. 1987a; Jackson et al. 1993; Marshall et al. 1995; Christopher et al. 2005; Montero-Casta{\~n}o et al. 2009; Liu et al. 2012; Mart{\'{\i}}n et al. 2012) surrounding the central black hole.
These gas structures are the most extreme environments for high-mass star formation in the Milky Way (Morris 1993 and references therein). 

The thermal dust continuum emission is the least biased by the chemical abundances and the excitation conditions. Hence, it is the most reliable tracer for demonstrating the reality and significance of the individual gas clouds or streamers, as well as their internal structures. 
The IRAM-30m and the JCMT single dish telescope observations of the millimeter and the submillimeter continuum emission have been presented by Zylka \& Mezger (1988), Mezger et al. (1989), Dent et al. (1993), Lis \& Carlstrom (1994), Pierce-Price et al. (2000), and Garc{\'{\i}}a-Mar{\'{\i}}n et al. (2011).
In this work, we report the first wide-field interferometric mosaic observations (157 pointings, $\sim$5$'$$\times$5$'$ field of view) of the 0.86 mm dust continuum emission using the Submillimeter Array  (SMA\footnote{The Submillimeter Array is a joint project between the Smithsonian Astrophysical Observatory and the Academia Sinica Institute of Astronomy and Astrophysics, and is funded by the Smithsonian Institution and the Academia Sinica.}; Ho et al. 2004).
The improved angular resolution of $\sim$5$''$ permits the detection of 0.1-0.2 parsec scale gas clumps, which can be the candidates of high-mass star-forming cores. 
In addition, we compare the dust continuum image with the HCO$^{+}$ 4-3 line image simultaneously obtained with the SMA.  We also compare these maps with the National Radio Astronomy Observatory (NRAO\footnote{The National Radio Astronomy Observatory is a facility of the National Science Foundation operated under cooperative agreement by Associated Universities, Inc.}) Robert C. Byrd Green Bank Telescope (GBT) observations of the CS/C$^{34}$S/$^{13}$CS 1-0 lines and the SiO 1-0 lines.  This allows us to examine the gravitational stabilities of the dense structures and the possible existence of shock fronts (e.g. Kauffmann et al. 2013). 
 
The new observations are described in Section \ref{chap_obs}.
The results are presented in Section \ref{chap_result}.
We compare our observations with the previous observations of OH, CH$_{3}$OH, and H$_{2}$O masers in Section \ref{chap_discussion}.
 A discussion of the non uniformity of the CND based on the 0.86 mm dust continuum image is given in Section \ref{sub_cnd}.
A brief summary of our results is given in Section \ref{chap_summary}.



\section{Observations and Data Reduction} 
\label{chap_obs}


\subsection{Submillimeter Continuum Emission}
\label{sub_smadata}
\subsubsection{ SMA Observations}
We made mosaic observations toward the Galactic center using the SMA, in its compact and subcompact array configurations.
These observations covered the frequency range of 354.1-358.1 GHz in the upper sideband, and 342.1-346.1 GHz in the lower sideband. 
Details of the subcompact array observations, and the pointing centers of the observations, can be found in Liu et al. (2012).
The compact array observations were made in two observing runs, on 2012 May 07 and 2012 May 20, with 6 and 7 available antennas, respectively.
The system temperatures T$_{sys}$ during these two runs were $\sim$180-400 K.
We observed two phase calibrators, 1733-130 and 1924-292, every $\sim$15 minutes during all observations. 
The amplitude and passband calibrators were Titan and 3C 279 on May 07, Neptune and 3C 279 on May 20. 
The target loops iterated over 157 pointing centers, with 5 5-second integrations at each pointing center. 
Each of the 157 pointings was visited more than three times in each observing run (i.e. on source time $>$3 hours in total in each run).
The minimum and maximum projected baselines in our SMA observations are $\sim$7.0 $k\lambda$ and $\sim$82 $k\lambda$.  
All SMA data were calibrated using the MIR IDL software package (Qi 2003). 
We used the MIRIAD (Sault 1995) task \texttt{UVAVER} to average all line-free channel data and reconstruct the 0.86 mm continuum band data.
The short spacing data were complemented by combining the archival JCMT SCUBA image (Appendix A, B).
The zeroth order free-free continuum emission model was constructed based on the archival VLA 7 mm observation data, and was subtracted from the combined SMA+JCMT 0.86 mm continuum image (Appendix C).
The simultaneously observed HCO$^{+}$ 4-3 line was regrided to 2.1 km\,s$^{-1}$ velocity channels for an adequate sensitivity, and will be presented for the purpose of discussing the virial condition.
The detailed studies of the submillimeter line data are deferred.


\subsection{GBT Observations}
\label{sub_nh3data}
We observed the CS 1-0 (48.99095 GHz) and the C$^{34}$S 1-0 (48.20692 GHz) transitions using the NRAO GBT on 2011 November 04 and 07. 
We observed the $^{13}$CS 1-0 (46.24754 GHz) and the SiO 1-0 (43.42376 GHz) transitions on 2011 November 07 and 09. 
The field of view of the SiO and $^{13}$CS observations is slightly offset toward the west to better recover the western gas streamers.
The angular resolution of the GBT is 763.8$''$/$\nu$, where $\nu$ is the observing frequency in GHz. 
The bright point source 1733-130 was observed in the beginning of each session for antenna surface (i.e. by Out of Focus Holography) and pointing calibrations. 
A line-free reference position RA: 17$^{\mbox{h}}$43$^{\mbox{m}}$43$^{\mbox{s}}$.344, Decl.: -29$^{\circ}$59$'$32$''$.27 was integrated for 30 seconds before and after the target observations in each block for off-source calibration data. 
We used the GBTIDL software package (Marganian et al. 2006) to calibrate the GBT data. 
We note that the CS isotopologue lines are close to the band edge of the GBT, which are subjected to the higher system temperature. 
We used the AIPS software package to perform imaging. 
We smoothed the final CS and C$^{34}$S image to an optimized $\theta_{maj}$$\times$$\theta_{min}$ = 20$''$$\times$18$''$, and BPA. = 0$^{\circ}$ to suppress the striping defects due to the sampling rates, and removed some obvious stripes by fitting zeroth order polynomial. 
The achieved RMS noise levels in each 24 kHz ($\sim$0.15 km\,s$^{-1}$ at 48.99 GHz) spectral channel are $\sim$0.5 K for the CS and C$^{34}$S observations, and are $\sim$0.14 K for the $^{13}$CS and SiO observations.



\section{Results}
\label{chap_result}
We present the observing results in this section. 
Our nomenclature follows Christopher et al. (2005), Amo-Baladr{\'o}n et al. (2011), and Liu et al. (2012). 

We also compare our results with the VLA observations of the 1612 MHz OH masers (Pihlstr\"{o}m et al. 2008; see also Sjouwerman 1998), the VLA observations of the 1720 MHz OH masers (Sjouwerman \& Pihlstr\"{o}m 2008), the GBT and VLA observations of the 44.1 GHz Class I  CH$_{3}$OH masers (Yusef-Zadeh 2008), the JVLA observations of the 44.1 GHz Class I CH$_{3}$OH masers (Pihlstr{\"o}m et al. 2011), the JVLA observations of the 36.2 GHz Class I CH$_{3}$OH masers (Sjouwerman 2010), and the GBT observations of the 22 GHz H$_{2}$O masers (Yusef-Zadeh 2008).

\begin{figure*}[h]
\vspace{-3cm}
\hspace{-1cm}
\includegraphics[width=22cm]{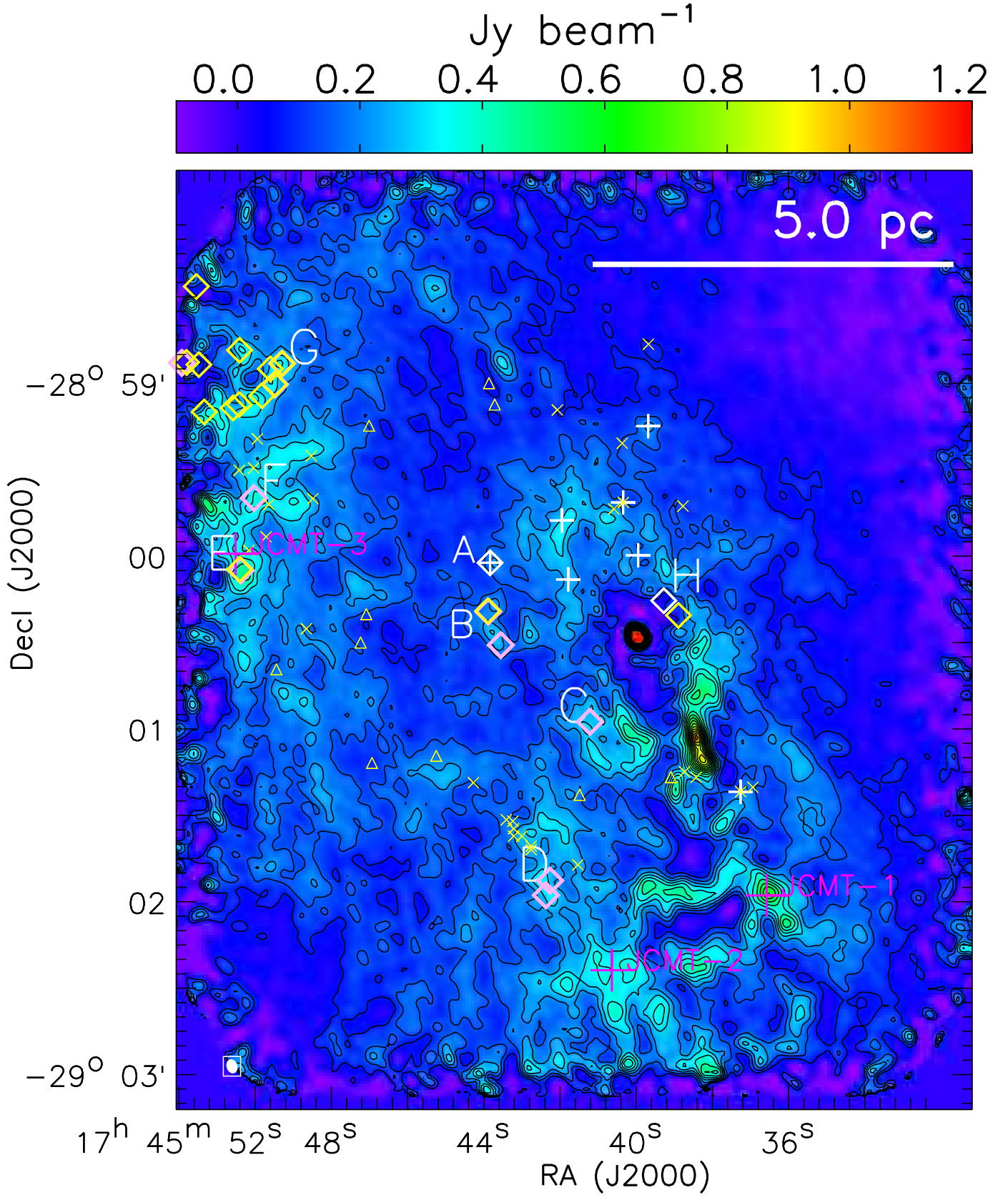} 
\vspace{-0.8cm}
\caption{\footnotesize{
The SMA+JCMT 0.86 mm continuum image with a free-free model subtracted. (color and contour). 
The angular resolution of this image is $\theta_{\mbox{\scriptsize{maj}}}\times\theta_{\mbox{\scriptsize{min}}}$=5$''$.1$\times$4$''$.2.
Contour spacings are 3$\sigma$ starting at 3$\sigma$ ($\sigma$=24 mJy\,beam$^{-1}$).
Yellow crosses are the 1720 MHz OH masers taken from Sjouwerman \& Pihlstr\"{o}m (2008).
Yellow triangles are the compact, either thermal or low-gain masing 1612 MHz OH line sources discussed in Pihlstr\"{o}m et al. (2008) (see also Sjouwerman 1998).
Pink diamonds are the 36.2 GHz Class I CH$_{3}$OH masers reported in Sjouwerman (2010).
Yellow diamonds are the 44.1 GHz Class I CH$_{3}$OH masers reported in Pihlstr\"{o}m et al. (2011).
White Crosses and diamonds are the 22 GHz water masers and the 44.1 GHz CH$_{3}$OH masers reported in Yusef-Zadeh (2008).
Note the symbol size is not representative to the spatial uncertainty.
For our convenience in discussion, we mark the peaks above the 45$\sigma$ significance level besides the Northeast Lobe and Southwest Lobe as JCMT-1,2,3 (see Appendix A).
The regions associated with CH$_{3}$OH masers are labeled by A-H.
}}
\label{fig_ch0ffsub}
\end{figure*}


\subsection{Continuum Emission}
\label{sub_ch0}
\begin{figure}[h]
\hspace{-1cm}
\begin{tabular}{c}
\vspace{-2cm} \\
\includegraphics[width=10.5cm]{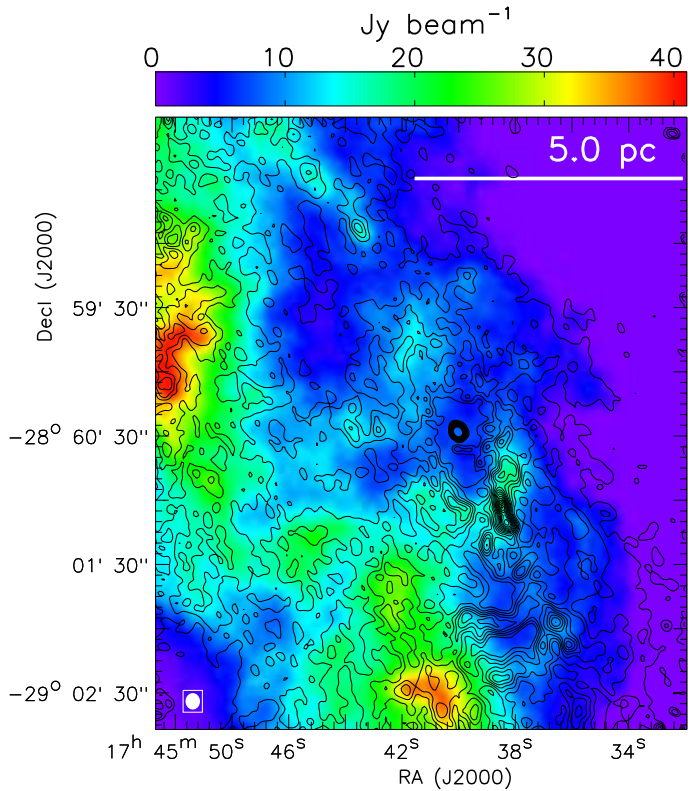} \\
\end{tabular}
\vspace{-0.4cm}
\caption{\footnotesize{
The JCMT SCUBA 0.44 mm (678 GHz) continuum image (color; $\theta_{\mbox{\scriptsize{maj}}}\times\theta_{\mbox{\scriptsize{min}}}$=8$''$$\times$8$''$), overlaid with the high angular resolution SMA+JCMT 0.86 mm continuum image (contour; $\theta_{\mbox{\scriptsize{maj}}}\times\theta_{\mbox{\scriptsize{min}}}$=5$''$.1$\times$4$''$.2).
Contour spacings are 3$\sigma$ starting at 3$\sigma$ ($\sigma$=72 mJy\,beam$^{-1}$).
The public processed JCMT SCUBA 0.44 mm was retrieved from the online data archive (ProjectID: M98AU64), and was published by Pierce-Price et al. (2000).
The beam of the JCMT SCUBA 0.44 mm observation is shown in the lower left. 
}}
\label{fig_jcmt690}
\end{figure}

Figure \ref{fig_ch0ffsub} shows the high angular resolution SMA+JCMT 0.86 mm continuum image.
A model of the free-free emission has been subtracted (see Appendix \ref{subsub_freefree}). 
The 0.86 mm continuum image recovers the detailed gas structures in the CND including the Northeast Lobe, the Northeast Extension, the Southwest Lobe, the Southern Extension, and the W-2,3,4 Streamers which connect to the CND from the west. 
It addition, it presents the detailed structures embedded in the Northern Ridge, the Southern Ridge, the 50 km\,s$^{-1}$ cloud, the Molecular Ridge, the 20 km\,s$^{-1}$ cloud. 
A southern dust ridge which appears to connect JCMT-1 to JCMT-2 (see Figure \ref{fig_ch0ffsub} ) in the archival JCMT SCUBA 0.44 mm (678 GHz) continuum image published by Pierce-Price et al. (2000; ProjectID: M98AU64), is also resolved in our SMA+JCMT image (see Figure \ref{fig_jcmt690}).
The spatial location of this southern dust ridge coincides with the Southern Arc reported by the previous observations of the CS 1-0 line emission (Liu et al. 2012).

Excluding the central point source, the 0.86 mm flux of the CND is approximately 68 Jy in a distribution that closely follows that of the inner 5 pc of  the CND.
At least few tens of dense molecular gas clumps are marginally resolved.
Higher angular resolution observations may resolve more blended dense clumps in this field. The precise number of clumps is not important for this discussion. The main conclusion is the impression of a very clumpy structure.   
The most significant dense clumps are found in the protrusion connecting to the Southwest Lobe.
Clusters of very dense clumps are also resolved towards the peaks of the JCMT SCUBA 0.86 mm image (e.g. JCMT-1,2,3 and the peaks in the Northern Ridge; see Appendix A).

\hspace{-1cm}
\begin{figure*}
\vspace{-1cm}
\begin{tabular}{c}
\\
\end{tabular}

\vspace{-1cm}
\begin{tabular}{ p{5.5cm} p{5.5cm} p{5.5cm} }
\includegraphics[width=6.5cm]{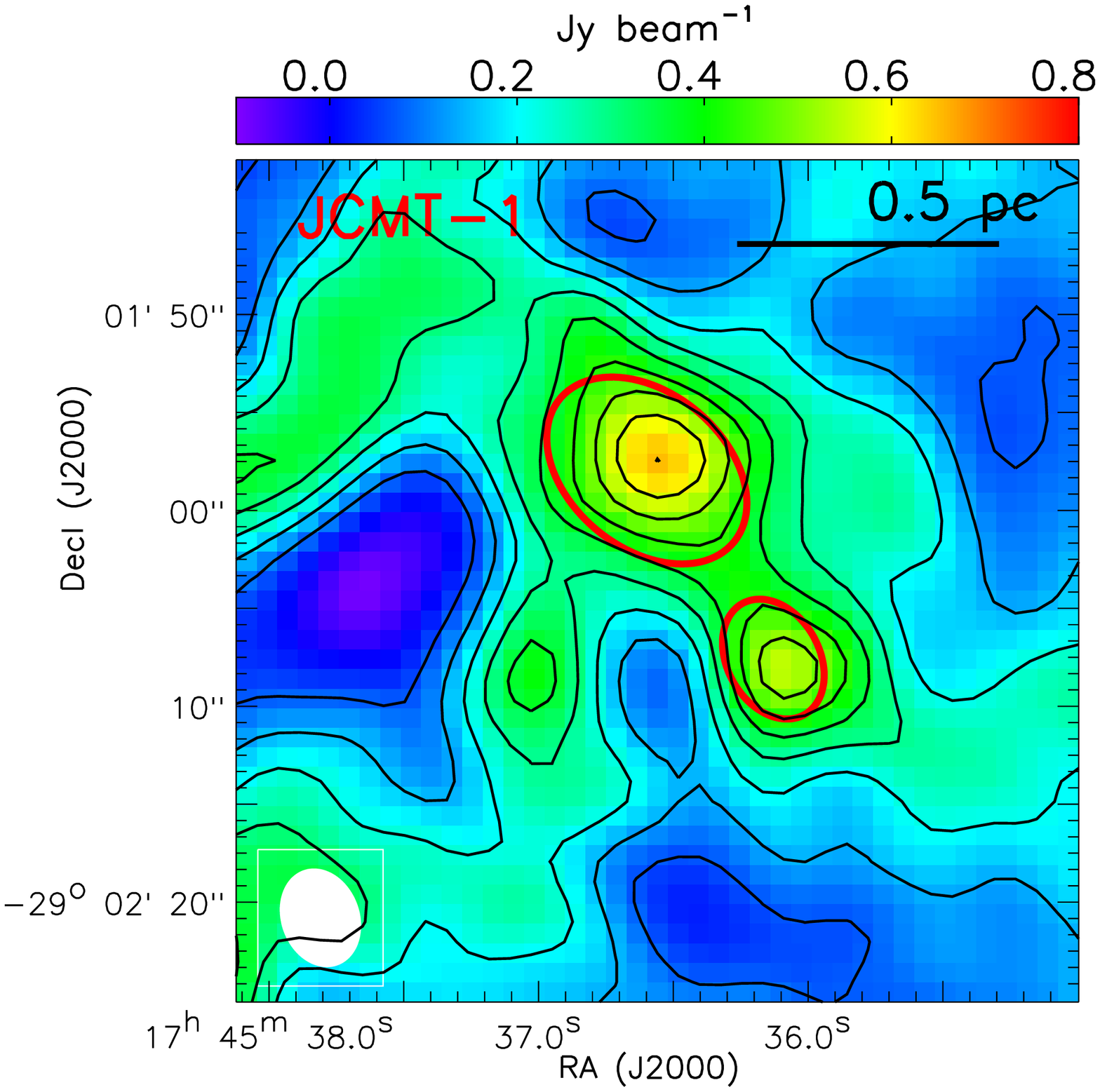} & \includegraphics[width=6.5cm]{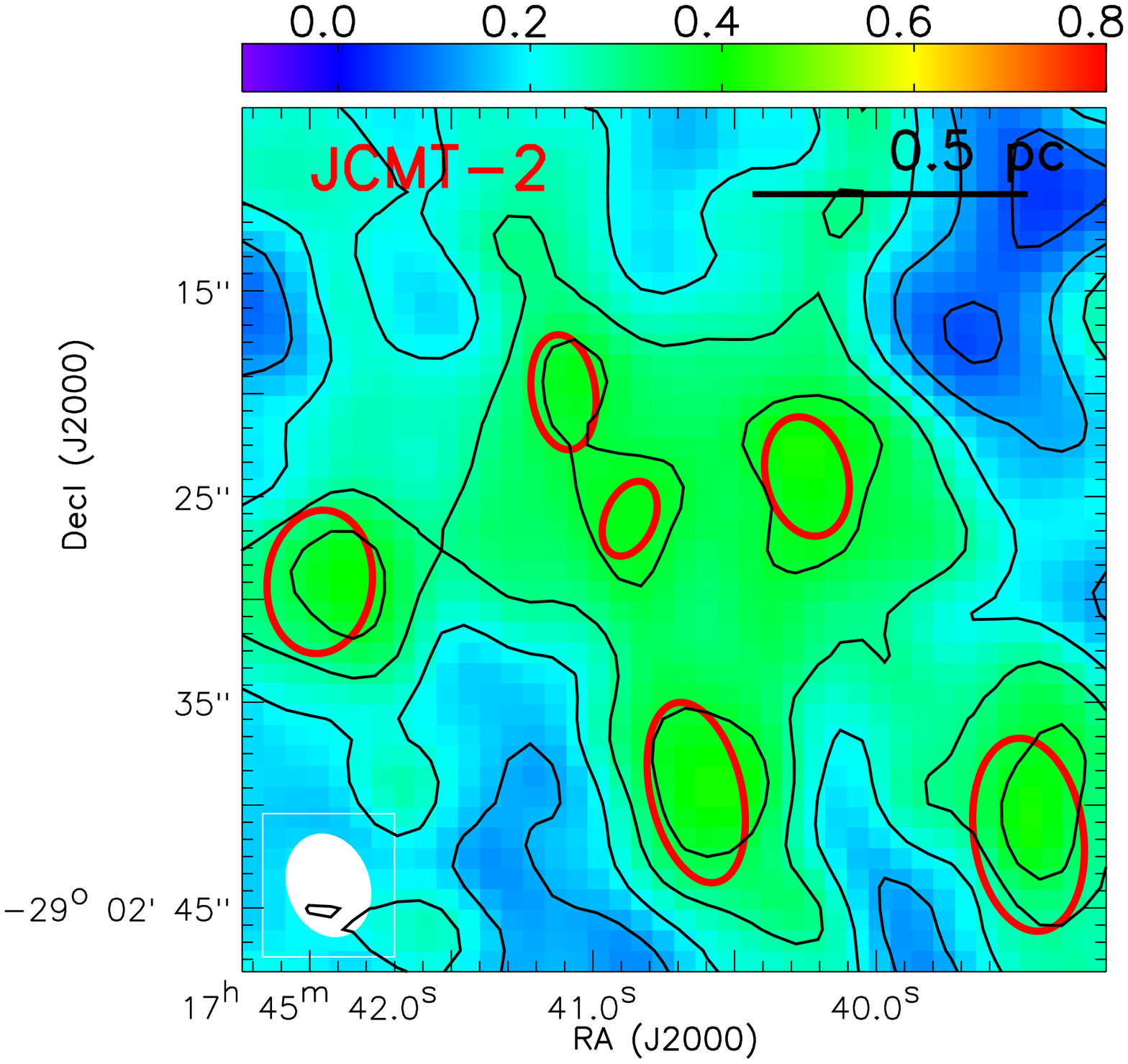} & \includegraphics[width=6.5cm]{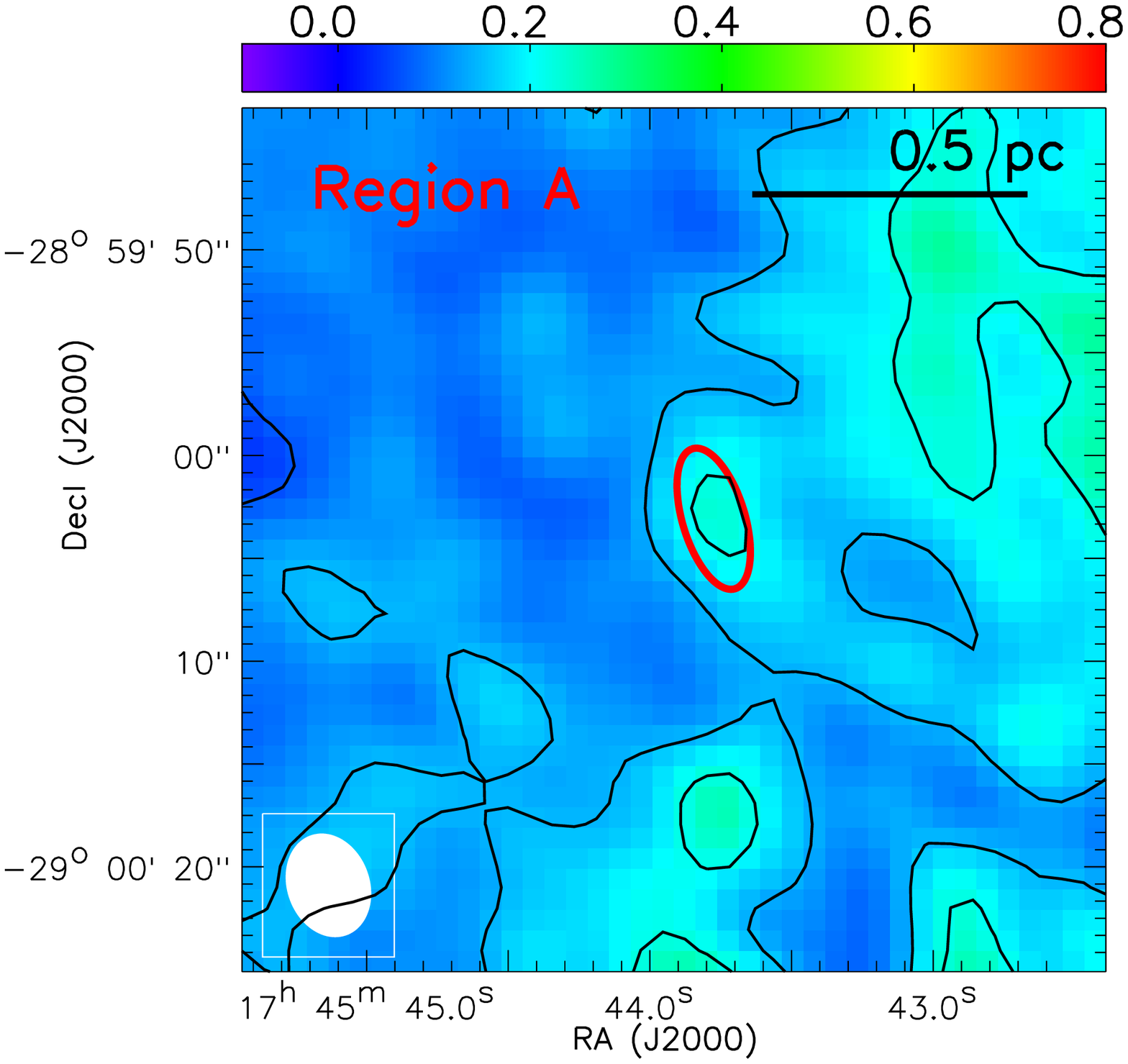} \\
\end{tabular}

\vspace{-2.2cm}

\begin{tabular}{ p{5.5cm} p{5.5cm} p{5.5cm} }
\includegraphics[width=6.5cm]{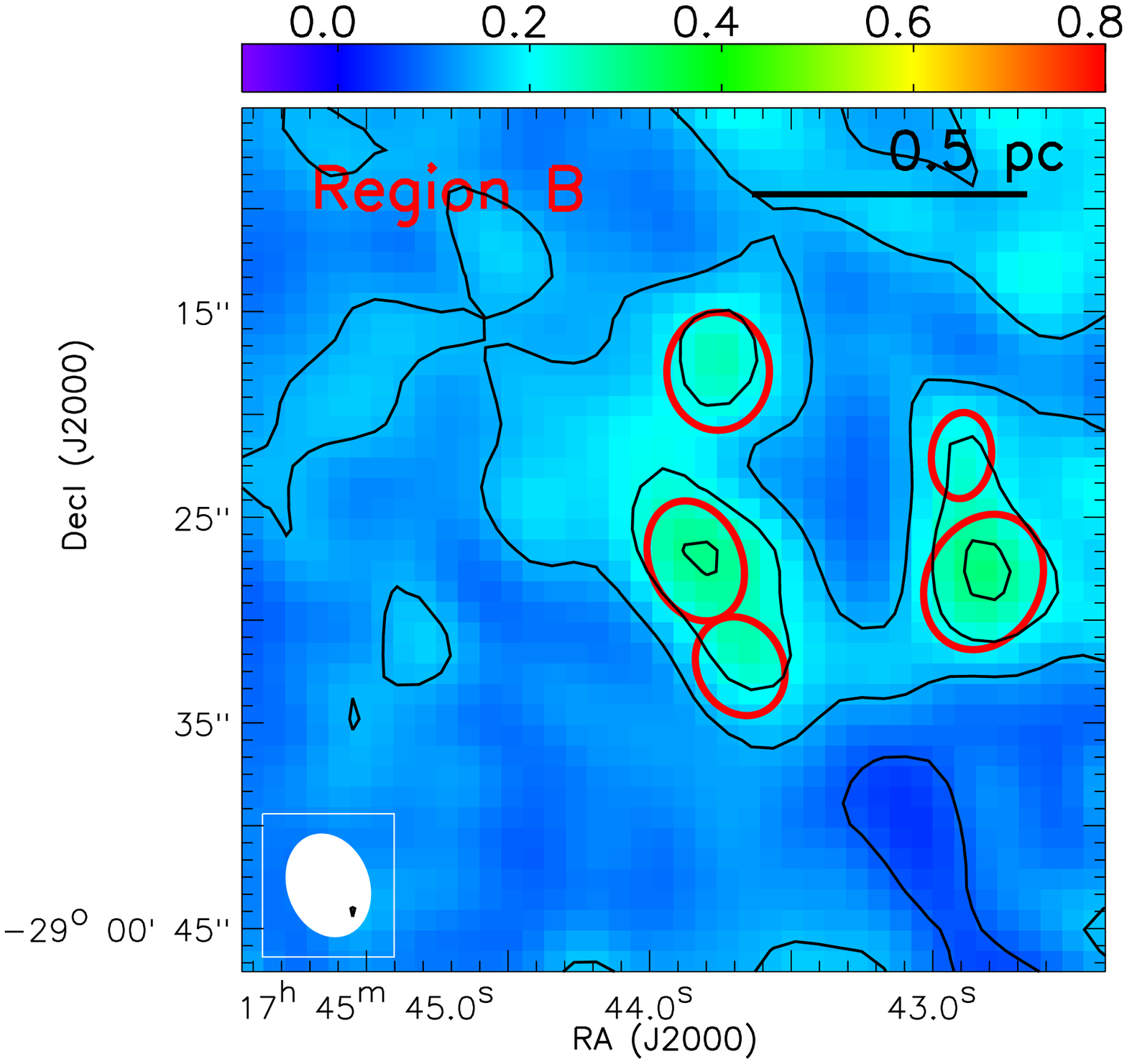} & \includegraphics[width=6.5cm]{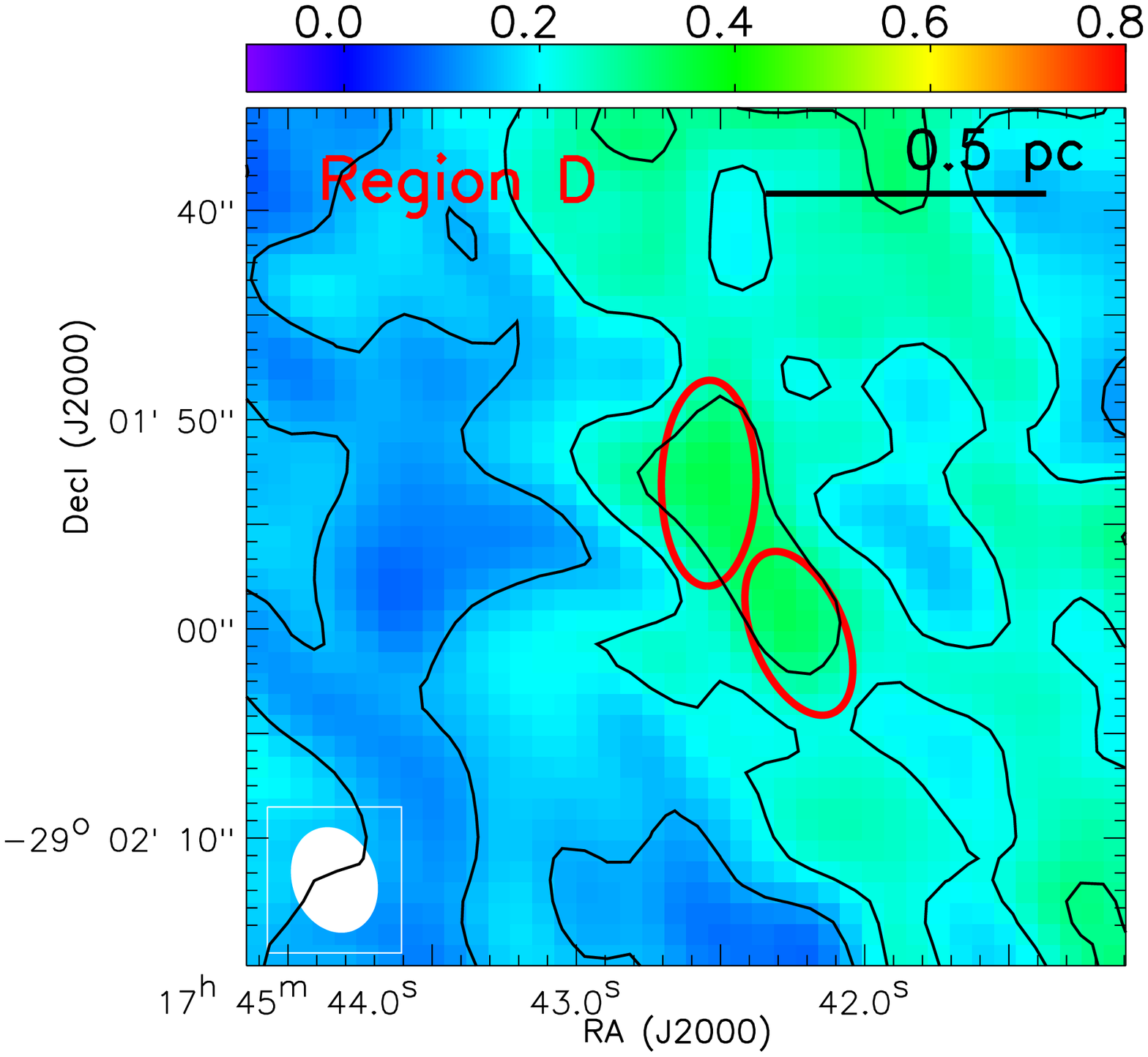} & \includegraphics[width=6.5cm]{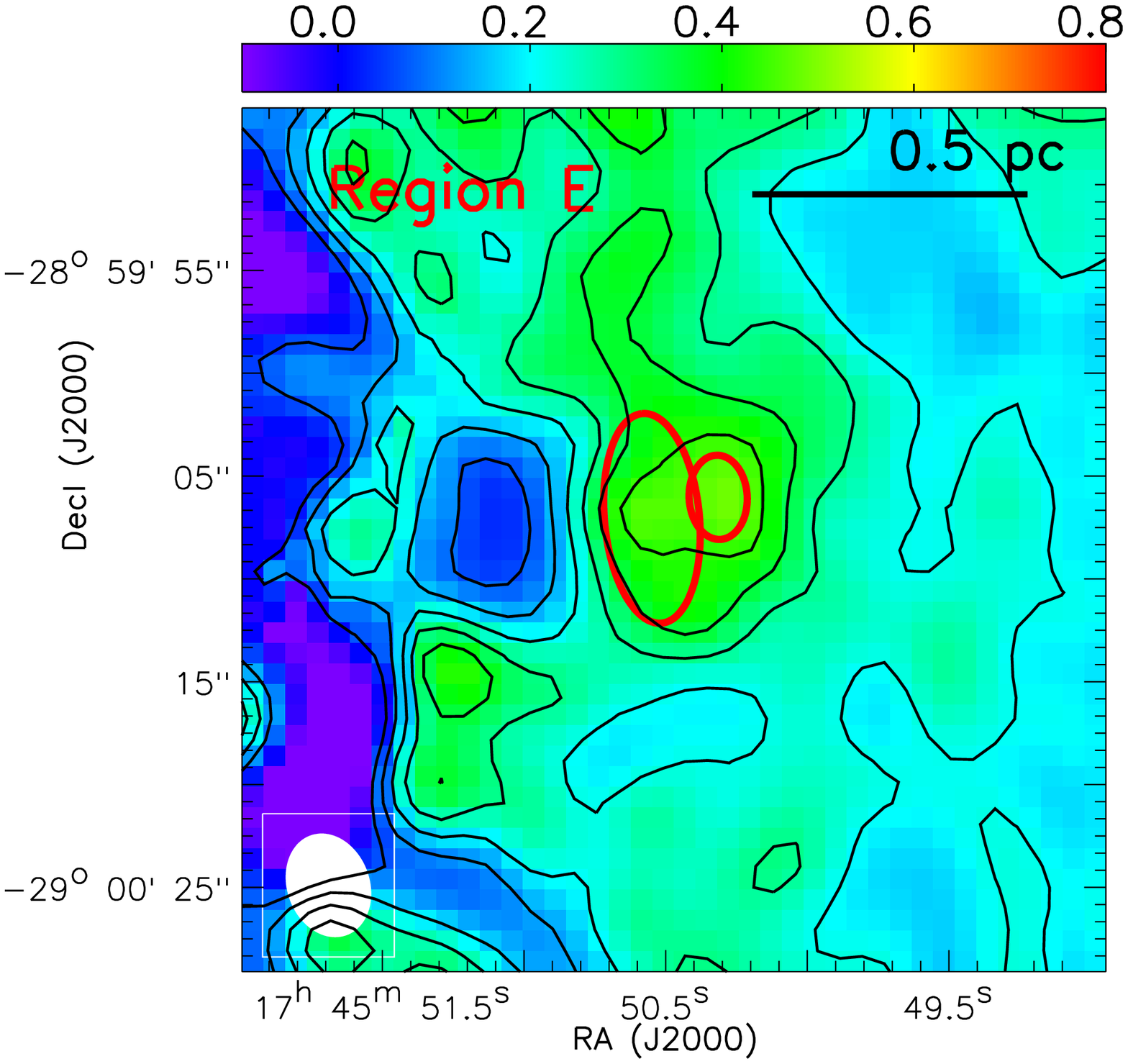} \\
\end{tabular}

\vspace{-2.2cm}

\begin{tabular}{ p{5.5cm} p{5.5cm} }
 \includegraphics[width=6.5cm]{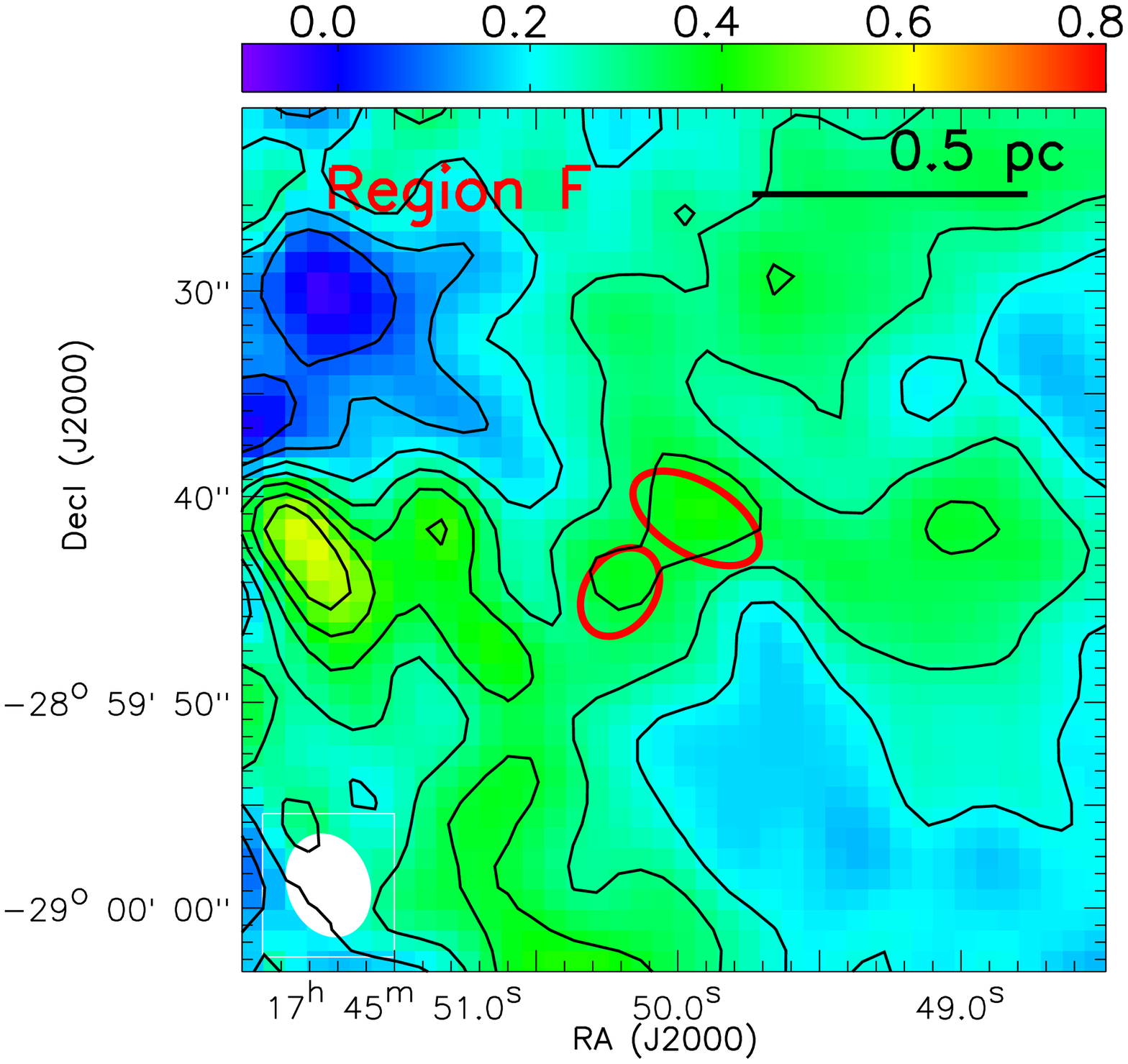} & \includegraphics[width=6.5cm]{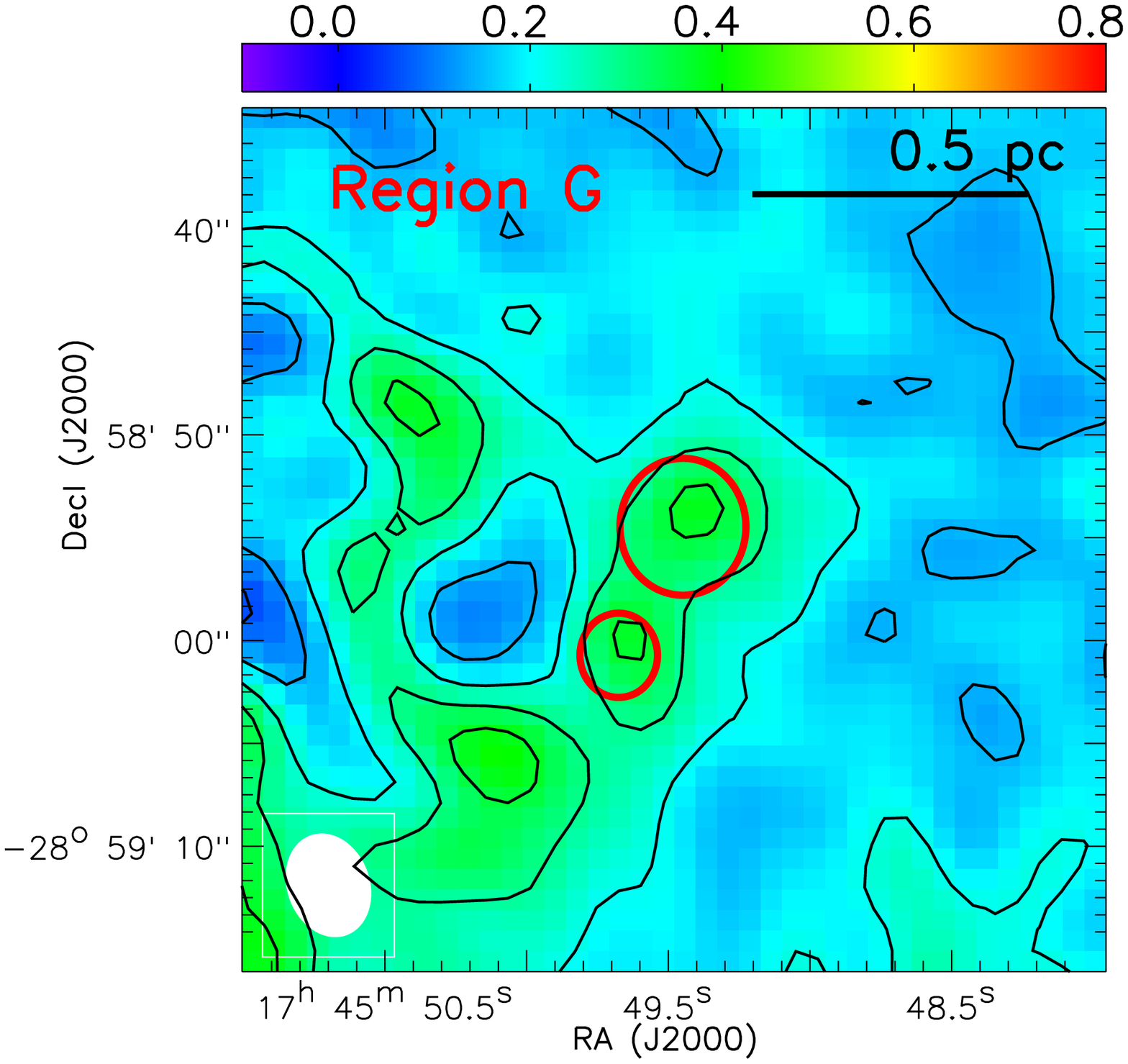} \\
\end{tabular}
\caption{\footnotesize{
Blowups of the free-free model subtracted SMA+JCMT 0.86 mm continuum image (color and contour) in parsec scale area around regions A, B, D, E, F, G, JCMT-1, and JCMT-2. 
Contour spacings are 3$\sigma$ starting at 3$\sigma$ ($\sigma$=24 mJy\,beam$^{-1}$).
Red ellipses show the fitted  2D Gaussian clumps. 
The major and minor axes of the red ellipses are the FWHM of the fitted 2D Gaussian (not deconvolved).
}}
\label{fig_regions}
\end{figure*}

In the present work, only the dense clumps located near the CH$_{3}$OH masers, and additionally the dense clumps located in JCMT-1 and JCMT-2 are discussed (Figure \ref{fig_regions}).
We do not attempt to systematically search for and analyze all dense clumps in the entire field because of the limited image quality, and insufficient angular resolution for the CND (see Montero-Casta{\~n}o et al. 2009 and Mart{\'{\i}}n et al. 2012 for higher angular resolution spectral line images). 
To estimate the 0.86 mm fluxes of the selected clumps, we first fit $\sim$1 pc scale 2D Gaussian components to suppress the contribution of the ambient gas, and then fit 2D Gaussian components to the localized peaks. 
The 0.86 mm fluxes of the individual Gaussian components are summarized in Table \ref{tab_clump}, which provides a quantitative measure of how massive the sub-parsec scale gas over-densities may be.
The uncertainties in background subtraction can give a few tens of percent of errors in fluxes.
In addition, the overestimates of the background emission can lead to underestimates of the clump sizescale.
The main purpose of this analysis is to estimate the virial velocity dispersions of these dense clumps (Section \ref{sub_clump}).
In this sense, the effects of overestimates of the background and underestimates of the clumps sizescale are competing. 
 Gaussian components with minor axis FWHM smaller than one standard deviation of the SMA+JCMT Gaussian beam width, 5$''$.2/2.355 = 2.2$''$ are not considered because we cannot estimate the physical deconvolved size scales.
The Gaussian components which are located near the edge of the SMA+JCMT 0.86 mm continuum image are also not considered.


\subsection{Molecular Emission}
\label{sub_line}
\begin{figure}
\begin{tabular}{c}
\includegraphics[width=9cm]{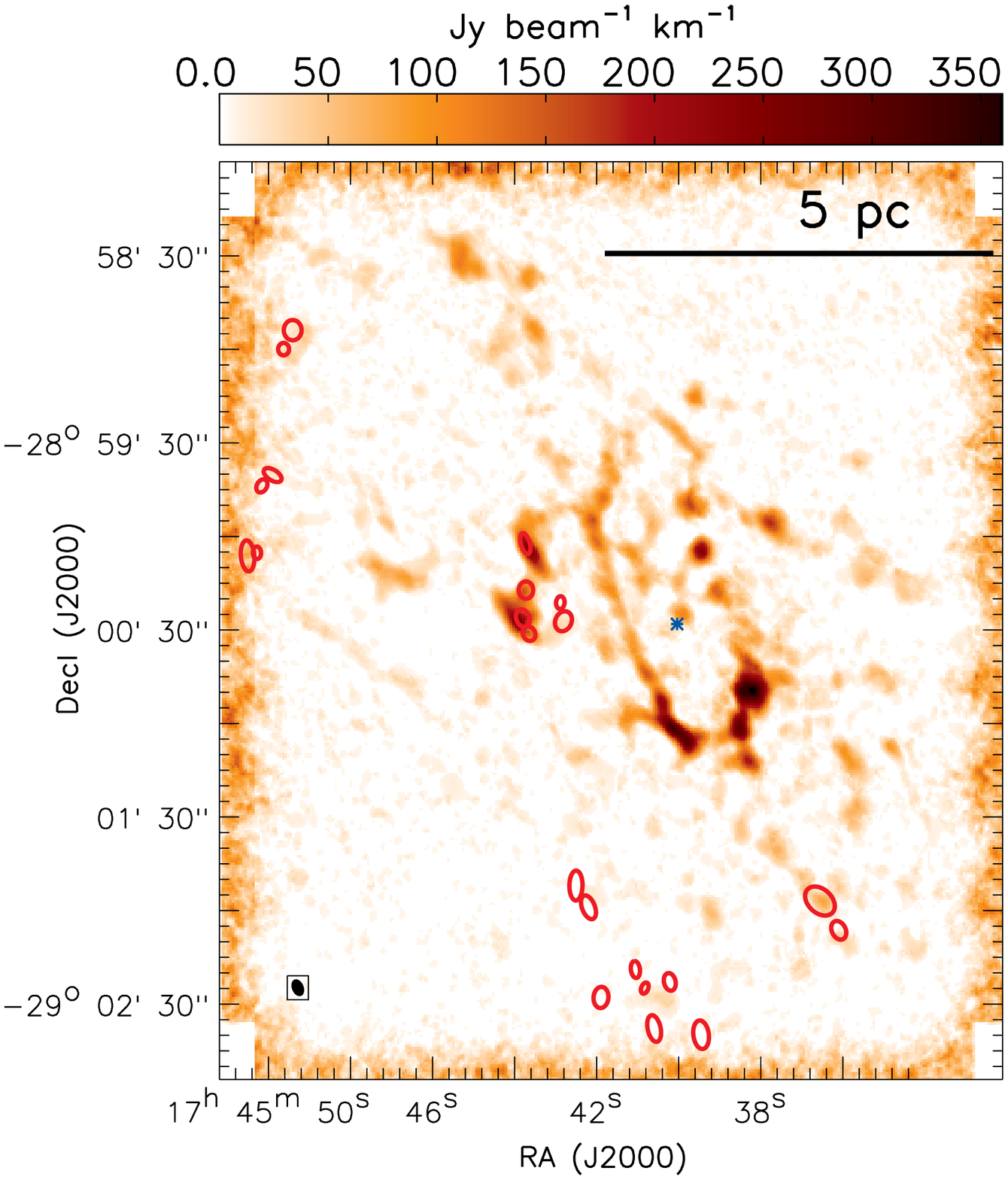} \\
\end{tabular}
\caption{\footnotesize{
The velocity integrated HCO$^{+}$ 4-3 image ($\theta_{\mbox{\scriptsize{maj}}}\times\theta_{\mbox{\scriptsize{min}}}$=5$''$.8$\times$4$''$.1).
Red ellipses show the fitted  0.86 mm clumps (see Figure \ref{fig_regions}). 
Blue star labels the location of Sgr A*.
}}
\vspace{0.3cm}
\label{fig_hcop}
\end{figure}

The SMA image of HCO$^{+}$ 4-3 is shown in Figure \ref{fig_hcop}, over-plotted with the fitted 0.86 mm dense clumps (Section \ref{sub_ch0}; Figure \ref{fig_regions}).
The HCO$^{+}$ 4-3 line is a good tracer of the ring-like CND, and also the gas streamers connecting to the CND. The eastern part of the CND
is relatively narrow and smooth,  compared to the western part.
The CND may be undergoing dynamical evolution, or is composed of distinct streams of molecular gas (Jackson et al. 1993; Wright et al. 2001; see Section \ref{sub_cnd}). 

\begin{figure}
\hspace{-0.5cm}
\includegraphics[width=9.5cm]{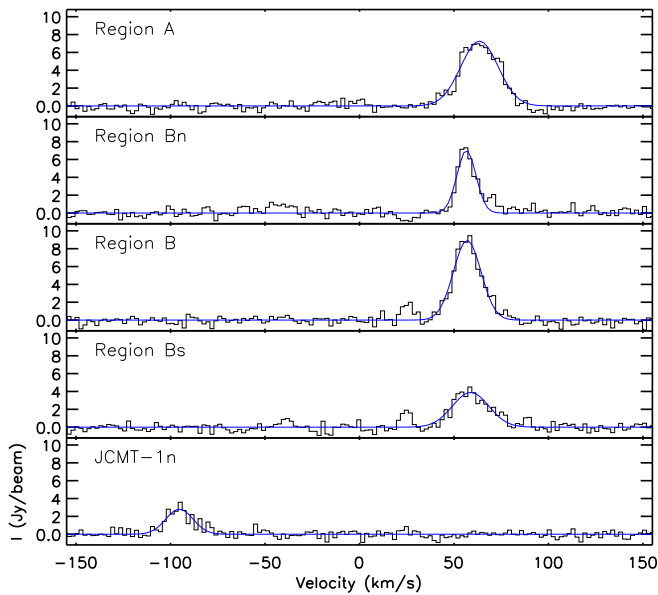}
\vspace{-1.3cm}
\caption{\footnotesize{
The HCO$^{+}$ 4-3 line spectra toward the selected regions (black; see Table \ref{tab_clump}). 
These spectra are regridded to 2.1 km\,s$^{-1}$ velocity channels.
The rms noise level is 0.6 Jy\,beam$^{-1}$ (1 Jy\,beam$^{-1}$$\sim$0.4 K). 
Blue lines present the fitted Gaussian components. 
}}
\vspace{0.3cm}
\label{fig_spectra}
\end{figure}

Towards the more extended 20 km\,s$^{-1}$ cloud, the Southern Arc, the Molecular Ridge and the 50 km\,s$^{-1}$ cloud, the SMA observations are detecting strong missing flux (see Liu et al. 2012).
In addition, the lower gas temperature in the extended cloud than in the CND also causes the weak or non-detection of the warm gas tracer HCO$^{+}$ 4-3 (E$_{\mbox{\scriptsize{up}}}$= 43 K). 
We therefore can only robustly image the HCO$^{+}$ 4-3 line towards a few fitted dense clumps.
The detected HCO$^{+}$ 4-3 spectra are presented in Figure \ref{fig_spectra}.
The velocity dispersions derived from the single component Gaussian fittings are summarized in Table \ref{tab_clump}.
From Figure \ref{fig_regions} and \ref{fig_spectra} we see that the brightest clump over the selected samples (Table \ref{tab_clump}), the JCMT-1n, has a centroid velocity of -95 km\,s$^{-1}$.
The JCMT-1n clump is likely to be embedded in the very blueshifted Southern Ridge, which shows strong SiO 2-1 emission in the earlier Nobeyama Millimetre Array observations (Sato \& Tsuboi 2008, and also see Amo-Baladr{\'o}n et al. 2011).
However, the CS 1-0 line spectrum taken at JCMT-1n indicates that a v$_{\mbox{\scriptsize{lsr}}}$$\sim$7 km\,s$^{-1}$ gas component is completely missed from the SMA observations of HCO$^{+}$ 4-3 (ref. spectrum 10 in Liu et al. 2012).
The virial velocity of JCMT-1n derived from the 0.86 mm dust continuum emission (Table \ref{tab_clump}) then should be considered as an upper limit when comparing with the HCO$^{+}$ 4-3 velocity dispersion. 

\begin{figure*}
\vspace{-2cm}
\begin{tabular}{c}
\\
\end{tabular}

\vspace{0cm}
\hspace{1.65cm}
\begin{tabular}{ p{6cm} p{6cm} }
\includegraphics[width=8.3cm]{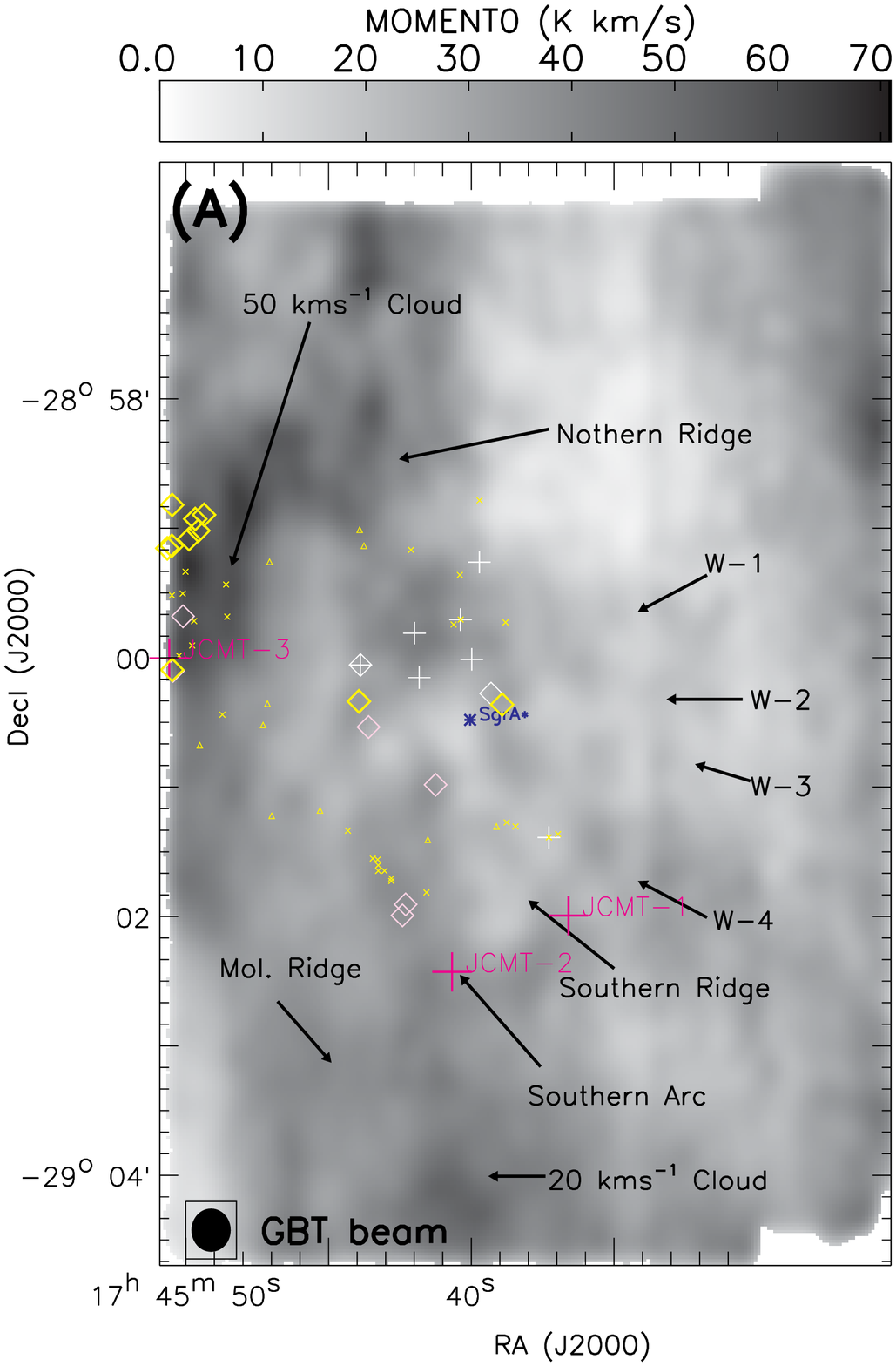} & \includegraphics[width=8.3cm]{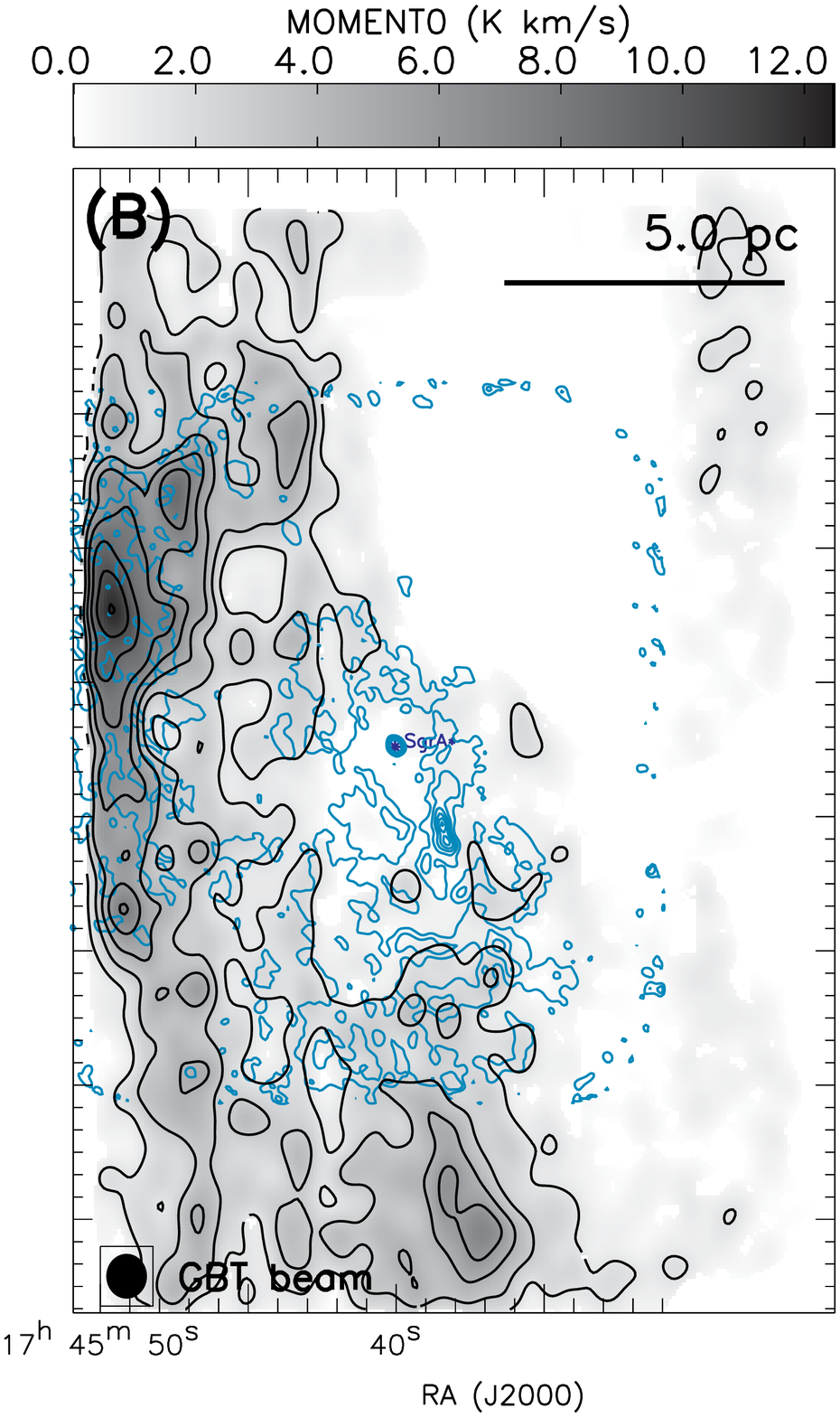}\\
\end{tabular}

\vspace{-2.2cm}
\hspace{1.65cm}
\begin{tabular}{ p{6cm} p{6cm} }
\includegraphics[width=8.3cm]{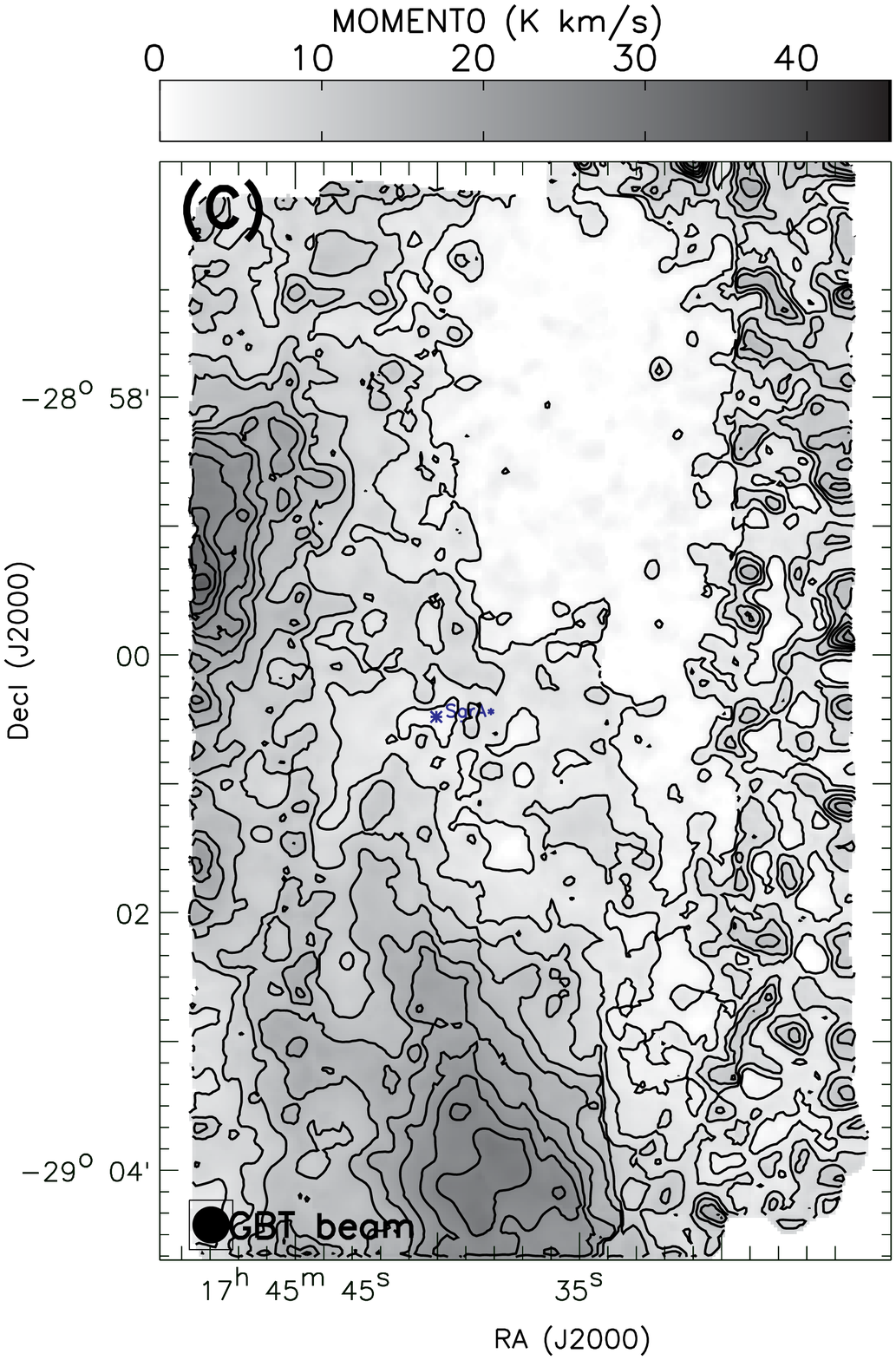} & \includegraphics[width=8.3cm]{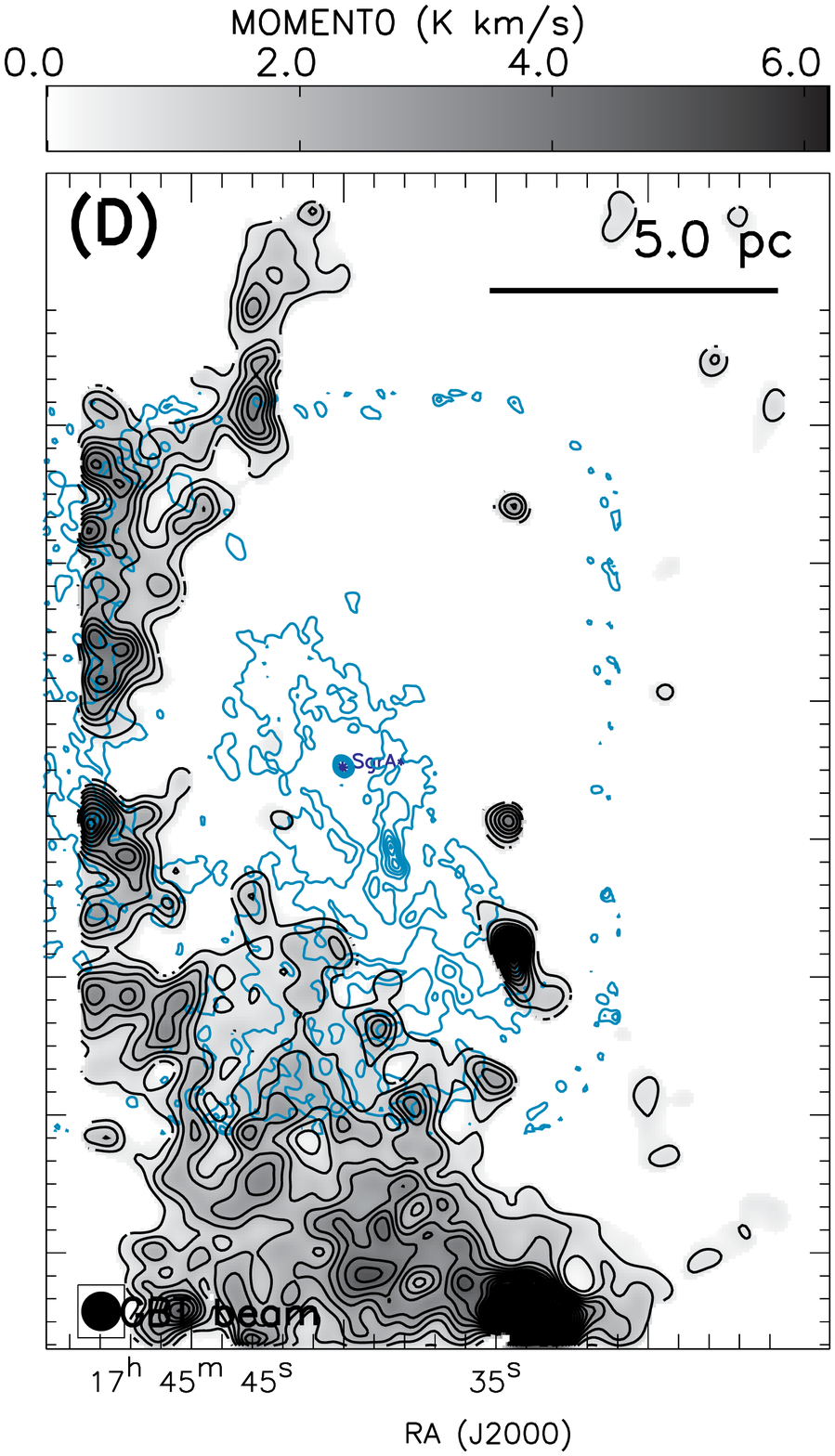}\\
\end{tabular}
\caption{\footnotesize{
Molecular gas in the central 20 pc region in the Galactic center. 
(A) The velocity integrated CS 1-0 emission. ($\sigma\sim$0.7 K\,km\,s$^{-1}$). The symbols are described in Figure \ref{fig_ch0ffsub}. Labels of the large scale gas streamers are consistent with those in Liu et al. (2012).
(B) The velocity integrated C$^{34}$S 1-0 emission (grayscale and black contour).  Contour spacings are 2$\sigma$ starting at 2$\sigma$ ($\sigma\sim$0.7 K\,km\,s$^{-1}$).
(C) The velocity integrated SiO 1-0 emission (grayscale and black contour). Contour spacings are 10$\sigma$ starting at 10$\sigma$ ($\sigma\sim$0.15 K\,km\,s$^{-1}$).
(D) The velocity integrated $^{13}$CS 1-0 emission (grayscale and black contour). Contour spacings are 4$\sigma$ starting at 4$\sigma$ ($\sigma\sim$0.15 K\,km\,s$^{-1}$).
The few $^{13}$CS peaks at R.A.=17$^{\mbox{\scriptsize{h}}}$45$^{\mbox{\scriptsize{m}}}$34-35$^{\mbox{\scriptsize{s}}}$ may be caused by ambiguities in baseline subtraction.
The rms noise level of the velocity integrated images are estimated based on the integration of the signal over a 13 km\,s$^{-1}$ velocity range, which is the median of the fitted velocity dispersion (see Table \ref{tab_clump}).
Green contours in panels (B) and (D) show the free-free model subtracted SMA+JCMT 0.86 mm continuum image.
The 0.86 mm continuum image contour spacings are 7.5$\sigma$ starting at 7.5$\sigma$ ($\sigma$=24 mJy\,beam$^{-1}$).
Blue star labels the location of Sgr A*.
}}
\label{fig_mnt0}
\end{figure*}

The more extended gas streamers are recovered by the GBT observations of the CS/C$^{34}$S/$^{13}$CS 1-0 lines and the SiO 1-0 line (Figure \ref{fig_mnt0}), which trace cooler gas (E$_{\mbox{\scriptsize{up}}}$= 2.1-2.3 K).
The SiO 1-0 emission appears to be correlated with the CS 1-0 emission. 
The flux ratio of these lines will be discussed in Section \ref{sub_ratio}.


\section{Discussion}
\label{chap_discussion}
We discuss the inferred ISM properties based on the presented observations. 

\begin{figure}
\vspace{-2.5cm}
\begin{tabular}{c}
\\
\end{tabular}

\hspace{-1cm}
\includegraphics[width=10.5cm]{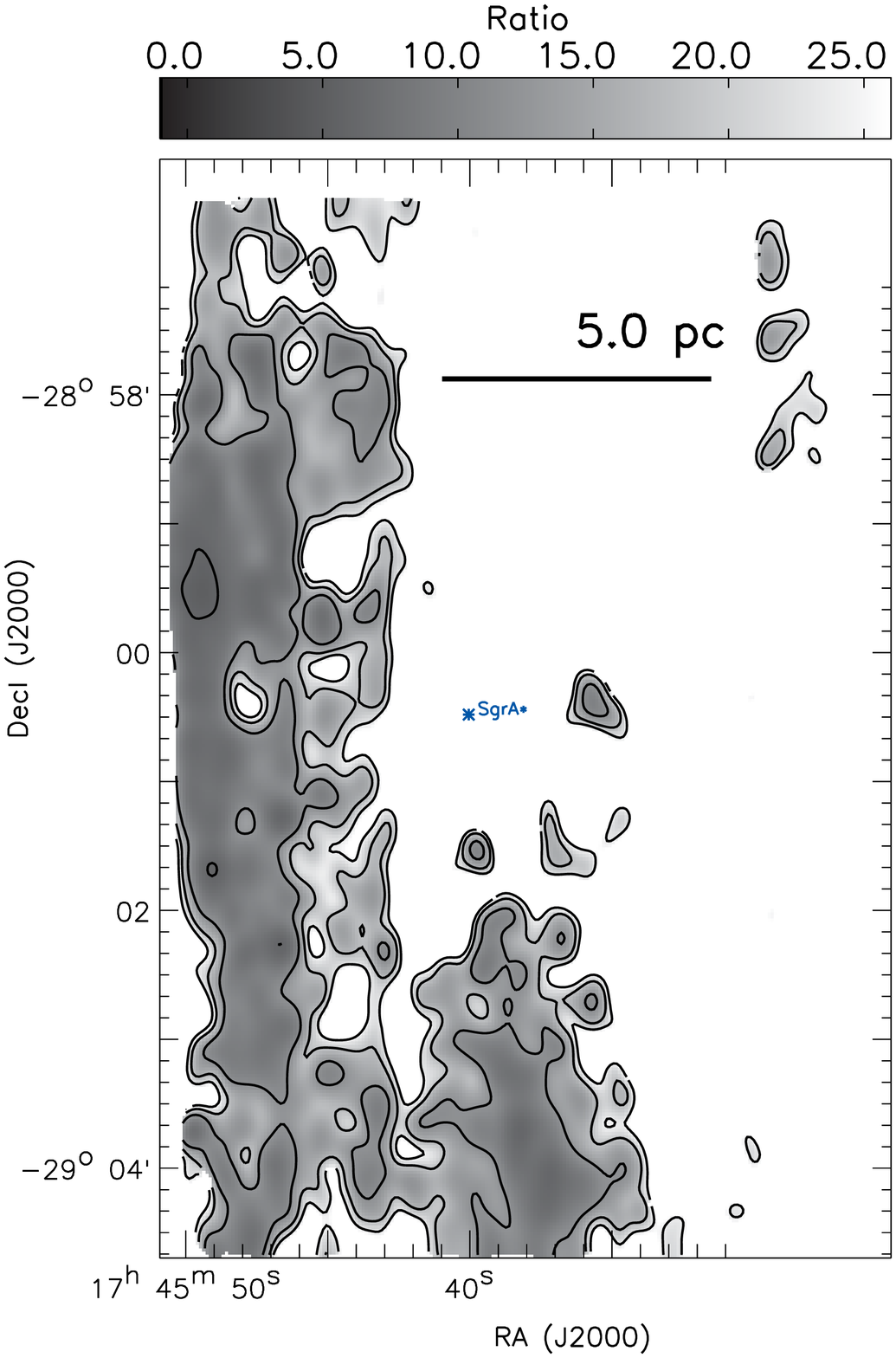}
\vspace{-0.6cm}
\caption{\footnotesize{
Velocity integrated CS/C$^{34}$S 1-0 ratio (color and contour).
Contour levels are [6, 12, 18, 24].
Blue star labels the location of Sgr A*.
}}
\label{fig_ratiocs}
\end{figure}


\subsection{The Spectral Line Ratio}
\label{sub_ratio}
While the abundance of the CS molecule is only mildly enhanced in UV and shocked environments (Amo-Baladr{\'o}n et al. 2011 and references therein), the abundance of the SiO molecule in shocked environments can be enhanced by up to a factor of 10$^{6}$ with respect to the value in quiescent gas (Martin-Pintado et al. 1992).
Since SiO and the CS isotopologues have similar dipole moments and energy level distributions, the derived flux ratio between these molecules can trace shocks without being sensitive to the assumed physical conditions.

Figure \ref{fig_ratiocs} shows the velocity integrated CS/C$^{34}$S 1-0 line ratio.
Above  $\sim$40 GHz,  difficulties in calibrating the several arcsecond GBT pointing offsets can potentially lead to the $\sim$30\% uncertainty in absolute flux level.
This uncertainty does not affect the derived CS/C$^{34}$S 1-0 ratio since these two lines are simultaneously observed. 
We found that the CS/C$^{34}$S 1-0 ratio is $\sim$6-12 towards the 50 km\,s$^{-1}$ cloud and the 20 km\,s$^{-1}$ cloud.
The earlier NRO 45m Telescope observations of CS 1-0 and C$^{34}$S 1-0 only robustly detected the C$^{34}$S 1-0 emission at the reference point located 3$'$ north and 3$'$ east of the Sgr A*, and showed a CS/C$^{34}$S 1-0 ratio of 8 (Tsuboi et al. 1999).
Considering that the NRO 45m Telescope observations were more beam smoothed, our results are consistent with the previous observations. 
Assuming the abundance ratio of [$X$(CS)/$X$(C$^{34}$S)] = 22.6 (Frerking 1980; Tsuboi et al. 1999), the optical depth $\tau$ of CS 1-0 can be estimated based on the following relation
\begin{equation}
\frac{S_{\mbox{\scriptsize{CS}}}}{S_{\mbox{\scriptsize{C$^{34}$S}}}}  = \frac{ 1 - \mbox{exp}(-\tau)  }{ 1 - \mbox{exp}(-\tau/22.6)  },
\end{equation}
where $S_{\mbox{\scriptsize{CS}}}$ and $S_{\mbox{\scriptsize{C$^{34}$S}}}$ are the fluxes of the CS and the C$^{34}$S lines.
The line ratio $S_{\mbox{\scriptsize{CS}}}/S_{\mbox{\scriptsize{C$^{34}$S}}}$=14.2 when $\tau$=1.
From Figure \ref{fig_ratiocs}, the majority of the optically thick CS 1-0 line ($\tau$=1.5-4, i.e. CS/C$^{34}$S 1-0 ratio $\sim$6-12) is seen towards the 50 km\,s$^{-1}$ cloud and the 20 km\,s$^{-1}$ cloud. 
The C$^{34}$S 1-0 line is optically thin over the observed field.

\begin{figure}
\vspace{-2.2cm}
\begin{tabular}{c}
\\
\end{tabular}
\hspace{-1cm}
\includegraphics[width=10.5cm]{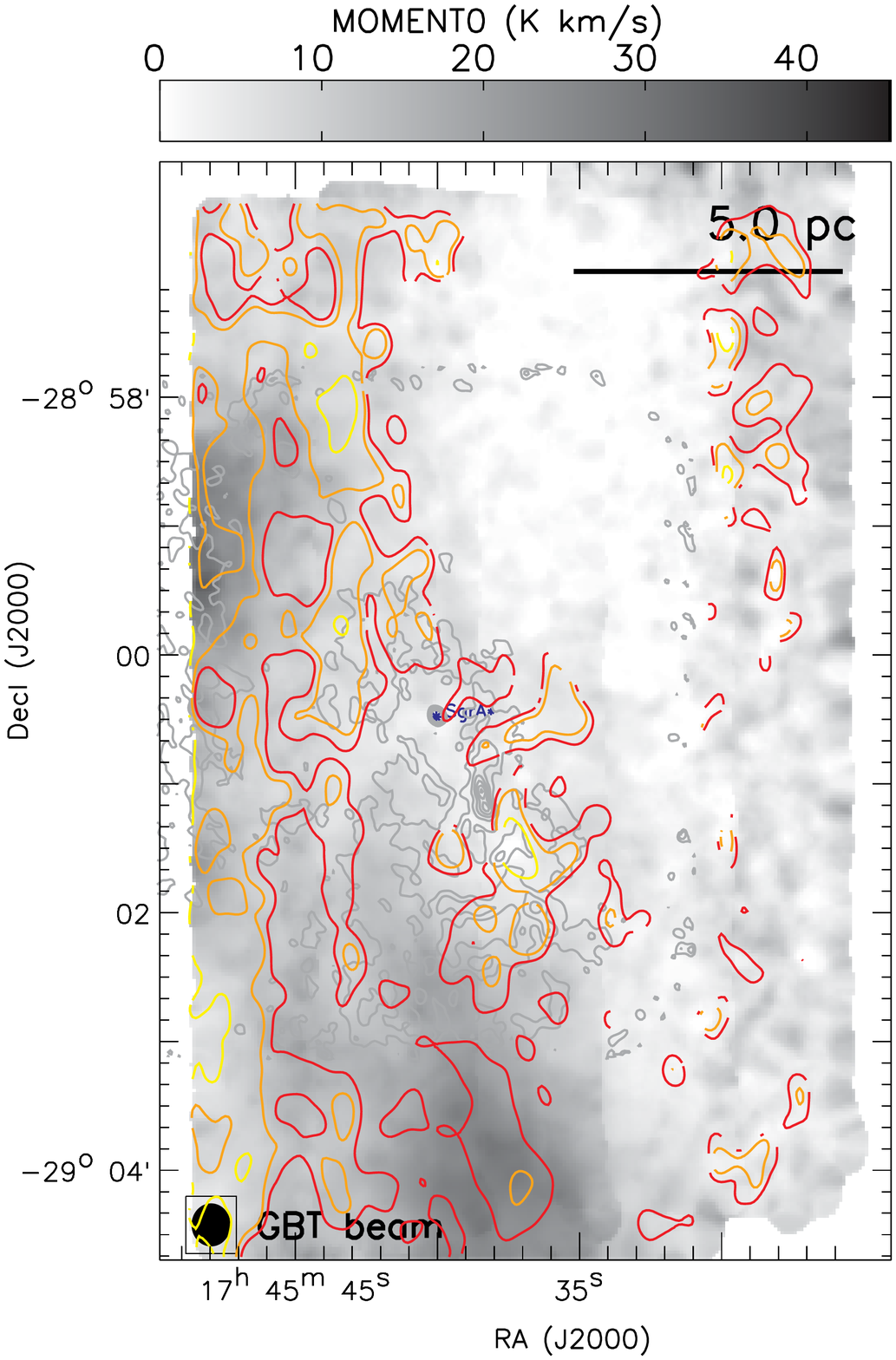}
\vspace{-0.6cm}
\caption{\footnotesize{
Velocity integrated SiO/C$^{34}$S 1-0 ratio (color contours).
The yellow, orange, and red contour levels are 2, 4, and 6, respectively. 
Gray contours show the free-free model subtracted SMA+JCMT 0.86 mm continuum image.
The velocity integrated SiO 1-0 emission is shown in grayscale.
The 0.86 mm continuum image contour spacings are 7.5$\sigma$ starting at 7.5$\sigma$ ($\sigma$=24 mJy\,beam$^{-1}$).
The GBT images are smoothed to the angular resolution of $\theta_{\mbox{\scriptsize{maj}}}\times\theta_{\mbox{\scriptsize{min}}}$=20$''$$\times$18$''$ before taking the ratio.
Blue star labels the location of Sgr A*.
}}
\label{fig_ratiosio}
\end{figure}

\begin{figure*}
\hspace{0cm}
\rotatebox{-90}{
\includegraphics[width=10.9cm]{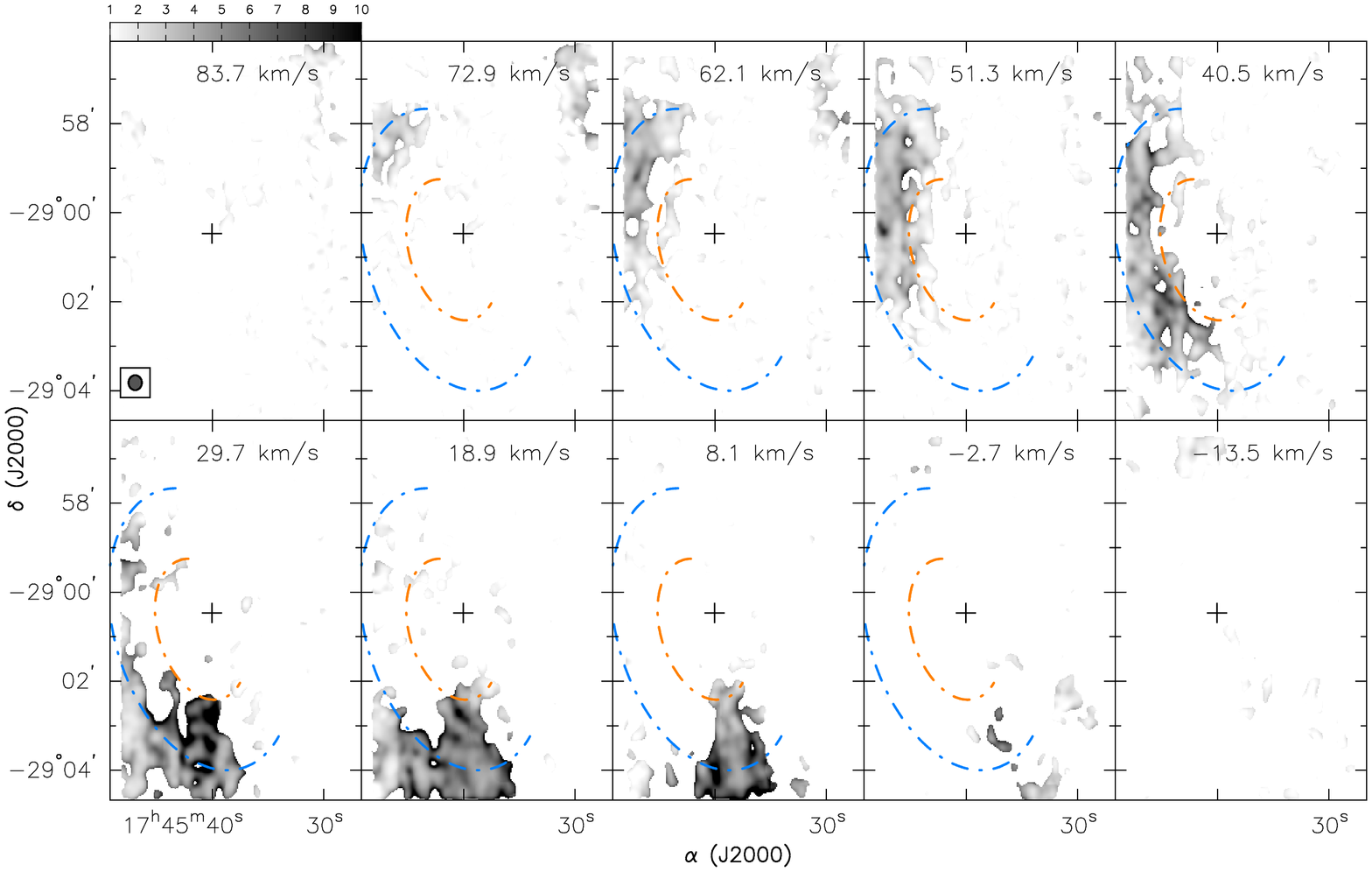}}
\caption{\footnotesize{
Velocity channel maps of the SiO/C$^{34}$S 1-0 ratio.
Plus signs label the location of Sgr A*.
The dash-dotted arcs are drawn to indicate the Southern Arc and the streamers in the east (consist of 50 km\,s$^{-1}$ cloud, the Molecular Ridge, and the 20 km\,s$^{-1}$ cloud).
The GBT images are smoothed to the angular resolution of $\theta_{\mbox{\scriptsize{maj}}}\times\theta_{\mbox{\scriptsize{min}}}$=20$''$$\times$18$''$ before taking the ratio.
}}
\vspace{0.3cm}
\label{fig_channel}
\end{figure*}

The SiO 1-0 line is $\sim$1.5-2 times fainter than the CS 1-0 line (Figure \ref{fig_mnt0}). 
Assuming the optically thin limit, the velocity integrated SiO/C$^{34}$S 1-0 line ratio (Figure \ref{fig_ratiosio})  traces the abundance ratio. 
We present the velocity channel maps of the SiO/C$^{34}$S 1-0 line ratio in Figure \ref{fig_channel} to trace the velocity structures of the shocked gas.
While the SiO/C$^{34}$S 1-0 ratio is already high in the entire map area, it appears to be enhanced in the 20 km\,s$^{-1}$ cloud and the Southern Arc.
Based on the correlation between NH$_{3}$ and 1.2 mm dust emission and the locations of OH masers, Wright et al. (2001) also suggested that these regions could be shock heated. 
Our observations do not have sufficient angular resolution to resolve the shock fronts.
Whether the shocks are created by the supernova shell to the south of the Sgr A East, or are created due to the non-circular orbits of the clouds, are not yet distinguished. 

The earlier IRAM-30 m Telescope observations additionally showed the enhanced SiO emission toward the 50 km\,s$^{-1}$ cloud and around the Southern Ridge and the W-4 Streamer. 
Our GBT observations of SiO 1-0 only covered a small part of the 50 km\,s$^{-1}$ cloud, and did not significantly detect the Southern Ridge.
The W-4 Streamer is only marginally detected in our C$^{34}$S 1-0 observations.


\subsection{The Molecular Gas Environment Around the Masers}
\label{sub_maserdust}
The 1720 MHz OH masers and the 1612 MHz OH masers trace the n$_{\mbox{\scriptsize{H}}_{2}}\sim$10$^{5}$ cm$^{-3}$ and n$_{\mbox{\scriptsize{H}}_{2}}\gtrsim$10$^{7}$ cm$^{-3}$ postshock gas in supernova remnants (assuming T$\sim$75 K; Pihlstr{\"o}m et al. 2008; Pavlakis \& Kylafis 1996).
The 22 GHz H$_{2}$O masers are collisionally excited at higher densities (n$_{\mbox{\scriptsize{H}}_{2}}\sim$10$^{7-9}$ cm$^{-3}$; Elitzur et al. 1992), and are often found in star-forming regions. 
The collisionally excited Class I CH$_{3}$OH masers are regarded as the unambiguous signposts of ongoing high-mass star formation in typical molecular clouds (Menten 1992; Yusef-Zadeh et al. 2008).
However, the Galactic center (e.g. inner 2-20 pc) gas streamers are warmer and 10 times more turbulent than the other star-forming molecular clouds.
In such extreme environments, the Class I CH$_{3}$OH masers may also be excited in cloud interactions although not yet observationally confirmed (Sjouwerman et al. 2010).

From Figure \ref{fig_jcmt}, we can see that on $\sim$10 pc scales, while the distribution of the OH masers follows a shell like geometry, the CH$_{3}$OH and H$_{2}$O masers are more scattered over the field. 
In the higher angular resolution dust continuum image (Figure \ref{fig_ch0ffsub}), we see that the CH$_{3}$OH and H$_{2}$O masers are preferentially detected near the $\lesssim$0.5 pc scale localized over-densities. 
The only exceptions are the CH$_{3}$OH masers associated with the region H which
 is located near the central bright free-free continuum sources.
The flux at region H may be biased by the uncertainties in interferometric imaging and the free-free model subtraction (Appendix \ref{subsub_freefree}).  
In addition, the molecular gas in region H may be photo-ionized by the central engine Sgr A*.

\begin{figure}[h]
\hspace{-1.3cm}
\includegraphics[width=10.5cm]{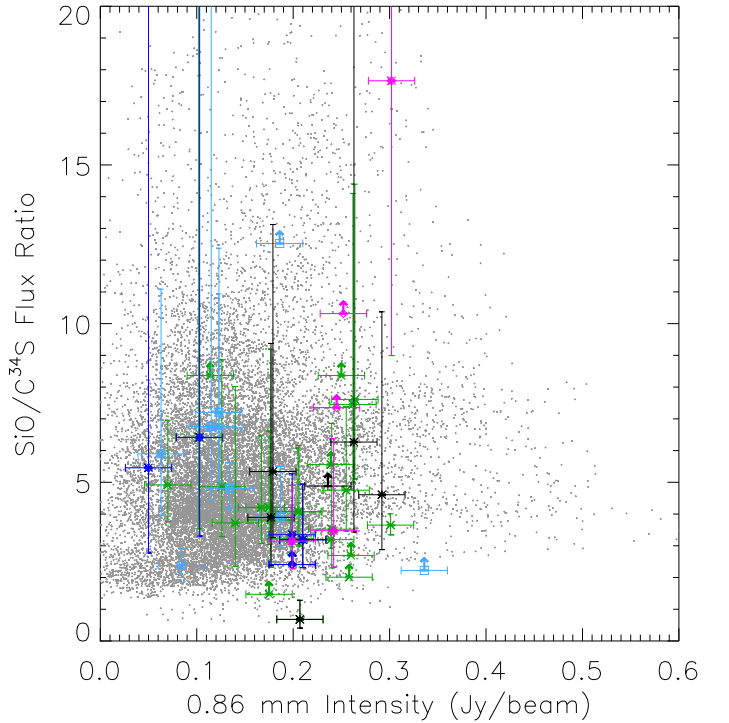}
\vspace{-0.6cm}
\caption{\footnotesize{
The velocity integrated SiO/C$^{34}$S 1-0 ratio and the 0.86 mm intensity at pixels where the velocity integrated C$^{34}$S 1-0 intensities are higher than 0.7 K\,km\,s$^{-1}$ (gray dots).
The values at the locations of the 1720 MHz OH masers (green), the 1612 MHz OH masers (light blue), the 22 GHz water masers (black), the 36.2 GHz CH$_{3}$OH masers (magenta), and the 44.1 GHz CH$_{3}$OH masers (blue) are presented by colored symbols.
For maser, the lower limit of the velocity integrated SiO/C$^{34}$S 1-0 ratio are given if the C$^{34}$S 1-0 emission is not robustly detected. 
The data points located outside of the field of view of either the 0.86 mm image or the SiO image are not presented. 
}}
\vspace{0.5cm}
\label{fig_diagram}
\end{figure}

We summarize the observed intensity of the dust continuum emission and the velocity integrated SiO/C$^{34}$S 1-0 ratio in Figure \ref{fig_diagram}.
Except for the 44.1 GHz CH$_{3}$OH masers associated with region H, the other CH$_{3}$OH and H$_{2}$O masers are detected at intensities $\gtrsim$0.15 Jy\,beam$^{-1}$.
The OH masers do not show a clear correlation with the flux of the dust continuum emission.
The clusters of OH maser may only sparsely sample the geometrically thin expanding shell (Coil \& Ho 1999) in projection of the bulk of the dense gas.
Yusef-Zadeh et al.(2008) and Pihlstr{\"o}m et al.(2011) only observed H$_{2}$O and CH$_{3}$OH masers towards selected high density regions, which potentially bias the corresponding maser data in Figure \ref{fig_diagram} toward higher `averaged' 0.86 mm intensity.
Nevertheless, the size scales of the dense clumps in Figure \ref{fig_ch0ffsub} are generally smaller than what can be discerned by previous observations of molecular gas.
We therefore do not think the spatial correlation between the clumps and the maser spots is purely an artifact due to the observational selection. 
As an example, the good correlation between the 0.86 mm emission clumps in region G (Figure \ref{fig_ch0ffsub}, i.e. pointing E in Pihlstr{\"o}m et al. 2011) and the JVLA detections of 44.1 GHz CH$_{3}$OH maser spots does not seem to be a coincidence, but should be interpreted with the local physical conditions.

For both the maser sources and the ambient gas, our observations do not resolve a clear correlation between the [$X$(SiO)/$X$(C$^{34}$S)] abundance ratio and the 0.86 mm intensity (Figure \ref{fig_diagram}).
On the large (e.g. $>$1 pc) scale, the shocked gas may be well mixed with the ambient material.
The enhanced abundance of SiO in shocks last for a few 10$^{3}$ years.
With the $\sim$50-100 km\,s$^{-1}$ relative motion of gas clouds around the CND, the smeared local SiO shock fronts may have widths of 0.15-0.3 pc (i.e. 3$''$.8-7$''$.5), which have to be examined with sensitive higher angular resolution observations. 
Here we refer to Mart{\'{\i}}n et al. (2012) for higher angular resolution observations of SiO emission in the 2$'$ field around the Sgr A*.
The velocity information is crucial for understanding whether and how the shock can induce the formation of sub-parsec scale dense gas structures.

\begin{table*}\scriptsize{
\vspace{1cm}
\caption{Properties of the fitted 22 0.86 mm continuum clumps.}
\label{tab_clump}
\hspace{-0.4cm}
\begin{tabular}{lcccccccccc}\hline\hline
Components	&	R.A. 			& Decl.		&	$\delta\theta^{\mbox{\scriptsize{maj}}}$	&	$\delta\theta^{\mbox{\scriptsize{min}}}$  & 0.86mm Flux	&	Gas Mass & Number Density	& $\delta v^{\mbox{\scriptsize{viriral}}}$	& $\delta v^{\mbox{\scriptsize{HCO$^{+}$}}}$	& $\delta v^{\mbox{\scriptsize{CS}}}$ \\
			&	(J2000)		& (J2000)	& (arcsec)	& (arcsec)	&	(mJy)	&	($M_{\odot}$)	&	(10$^{5}$ cm$^{-3}$) &	(km\,s$^{-1}$)	&	(km\,s$^{-1}$)	& (km\,s$^{-1}$)	\\\hline
Region A		& 17$^{\mbox{\scriptsize{h}}}$45$^{\mbox{\scriptsize{m}}}$43.74$^{\mbox{\scriptsize{s}}}$	& -29$^{\circ}$00$'$03$''$	&7.0	&	3.2	&	90	 &	13-31	&	2.5-5.9	&	0.4-0.8	&	9.7	& 19\\
Region Bn		& 17$^{\mbox{\scriptsize{h}}}$45$^{\mbox{\scriptsize{m}}}$43.72$^{\mbox{\scriptsize{s}}}$	& -29$^{\circ}$00$'$18$''$	&5.6	&	5.3	&	172	&	24-60	& 1.8-4.4	&	0.45-0.96	&	4.5	& 15\\
Region B				& 17$^{\mbox{\scriptsize{h}}}$45$^{\mbox{\scriptsize{m}}}$43.80$^{\mbox{\scriptsize{s}}}$	& -29$^{\circ}$00$'$27$''$	&6.0	&	4.7	&	238	&	34-83	& 2.9-7.2	&	0.56-1.1	&	7.0	& 14\\
Region Bs				& 17$^{\mbox{\scriptsize{h}}}$45$^{\mbox{\scriptsize{m}}}$43.65$^{\mbox{\scriptsize{s}}}$	& -29$^{\circ}$00$'$32$''$	&5.0	&	4.3	&	114	&	16-40	& 4.4-11	&	0.46-0.96	&	9.4	& 14\\
Region Bwn				& 17$^{\mbox{\scriptsize{h}}}$45$^{\mbox{\scriptsize{m}}}$42.89$^{\mbox{\scriptsize{s}}}$	& -29$^{\circ}$00$'$22$''$	&4.1	&	3.1	&	54	&	8-19	&	$\cdots$	& $\cdots$ &	$\cdots$	& 21\\
Region Bws			& 17$^{\mbox{\scriptsize{h}}}$45$^{\mbox{\scriptsize{m}}}$42.81$^{\mbox{\scriptsize{s}}}$	& -29$^{\circ}$00$'$28$''$	&6.9	&	5.7	&	317	&	45-110	& 1.5-3.6 &	0.55-1.1	&	$\cdots$	& 19\\
Region Dn	& 17$^{\mbox{\scriptsize{h}}}$45$^{\mbox{\scriptsize{m}}}$42.51$^{\mbox{\scriptsize{s}}}$	& -29$^{\circ}$01$'$53$''$	&9.6	&	4.8	&	271	&	38-94	& 0.88-2.1 &	0.47-0.96	&	$\cdots$	& 14\\
Region Ds	& 17$^{\mbox{\scriptsize{h}}}$45$^{\mbox{\scriptsize{m}}}$42.20$^{\mbox{\scriptsize{s}}}$	& -29$^{\circ}$02$'$00$''$	&8.2	&	4.6	&	195	&	28-68	& 1.0-2.6 &	0.43-0.87	&	$\cdots$	& 12\\
Region Ew	& 17$^{\mbox{\scriptsize{h}}}$45$^{\mbox{\scriptsize{m}}}$50.28$^{\mbox{\scriptsize{s}}}$	& -29$^{\circ}$00$'$06$''$	&4.0	&	3.0	&	121	&	17-42	&	$\cdots$ & $\cdots$	&	$\cdots$	& 13\\
Region Ee	& 17$^{\mbox{\scriptsize{h}}}$45$^{\mbox{\scriptsize{m}}}$50.51$^{\mbox{\scriptsize{s}}}$	& -29$^{\circ}$00$'$07$''$	&10.0	&	4.9	&	474	&	67-160	& 1.3-3.2 &	0.61-1.2	&	$\cdots$	& 13\\
Region Fw	& 17$^{\mbox{\scriptsize{h}}}$45$^{\mbox{\scriptsize{m}}}$49.90$^{\mbox{\scriptsize{s}}}$	& -28$^{\circ}$59$'$41$''$	&7.1	&	3.5	&	134	&	19-46 & 2.6-6.4	&	0.45-0.87	&	$\cdots$	& 12\\
Region Fe	& 17$^{\mbox{\scriptsize{h}}}$45$^{\mbox{\scriptsize{m}}}$50.16$^{\mbox{\scriptsize{s}}}$	& -28$^{\circ}$59$'$44$''$	&4.7	&	3.5	&	83	&	12-29	& 59-140 &	0.64-1.3	&	$\cdots$	& 12\\
Region Gn	& 17$^{\mbox{\scriptsize{h}}}$45$^{\mbox{\scriptsize{m}}}$49.41$^{\mbox{\scriptsize{s}}}$	& -28$^{\circ}$58$'$54$''$	&6.5	&	6.5	&	364	&	51-130	& 1.5-3.6 &	0.57-1.1	&	$\cdots$	& 12\\
Region Gs	& 17$^{\mbox{\scriptsize{h}}}$45$^{\mbox{\scriptsize{m}}}$49.63$^{\mbox{\scriptsize{s}}}$	& -28$^{\circ}$59$'$00$''$	&4.0	&	4.0	&	147	&	21-51	& 320-760 &	1.0-2.1	&	$\cdots$	& 13\\\hline
JCMT-1n	& 17$^{\mbox{\scriptsize{h}}}$45$^{\mbox{\scriptsize{m}}}$36.56$^{\mbox{\scriptsize{s}}}$	& -29$^{\circ}$01$'$58$''$	&11.9	&	8.0	&	1810	&	260-630 & 1.4-3.4	&	0.96-1.9	&	6.8	& 24\\
JCMT-1s	& 17$^{\mbox{\scriptsize{h}}}$45$^{\mbox{\scriptsize{m}}}$36.10$^{\mbox{\scriptsize{s}}}$	& -29$^{\circ}$02$'$07$''$	&6.5	&	4.9	&	527	&	74-180	& 4.4-11 &	0.77-1.6	&	$\cdots$	& 29\\
JCMT-2a	& 17$^{\mbox{\scriptsize{h}}}$45$^{\mbox{\scriptsize{m}}}$41.06$^{\mbox{\scriptsize{s}}}$	& -29$^{\circ}$02$'$20$''$	&5.5	&	3.3	&	119	&	17-41 & 16-39	&	0.58-1.1	&	$\cdots$	&  10 (or 19)\\
JCMT-2b	& 17$^{\mbox{\scriptsize{h}}}$45$^{\mbox{\scriptsize{m}}}$40.22$^{\mbox{\scriptsize{s}}}$	& -29$^{\circ}$02$'$24$''$	&5.8	&	4.2	&	169	&	24-59	& 3.6-8.8 &	0.51-1.0	&	$\cdots$	& 11 (or 20)\\
JCMT-2c	& 17$^{\mbox{\scriptsize{h}}}$45$^{\mbox{\scriptsize{m}}}$40.83$^{\mbox{\scriptsize{s}}}$	& -29$^{\circ}$02$'$26$''$	&3.8	&	2.5	&	48	&	6.6-16 &	$\cdots$ 	&	$\cdots$	&	$\cdots$	& 11 (or 18)\\
JCMT-2d	& 17$^{\mbox{\scriptsize{h}}}$45$^{\mbox{\scriptsize{m}}}$40.60$^{\mbox{\scriptsize{s}}}$	& -29$^{\circ}$02$'$39$''$	&8.8	&	4.7	&	412	&	58-140	&	1.8-4.1 & 0.61-1.2	&$\cdots$		& 11 (or 17)\\
JCMT-2e	& 17$^{\mbox{\scriptsize{h}}}$45$^{\mbox{\scriptsize{m}}}$41.90$^{\mbox{\scriptsize{s}}}$	& -29$^{\circ}$02$'$29$''$	&6.8	&	5.4	&	398	&	56-140	&	2.2-5.6	& 0.63-1.3	&	$\cdots$	& 10 (or 23)\\
JCMT-2f		& 17$^{\mbox{\scriptsize{h}}}$45$^{\mbox{\scriptsize{m}}}$39.46$^{\mbox{\scriptsize{s}}}$	& -29$^{\circ}$02$'$41$''$	&9.4	&	5.7	&	614	&	87-210	& 1.4-3.5 &	0.68-1.4	&$\cdots$		& 11 (or 26)\\\hline
\end{tabular}
}
\vspace{0.0cm}
\footnotesize{Note.---The $\delta\theta^{\mbox{\scriptsize{maj}}}$ and $\delta\theta^{\mbox{\scriptsize{min}}}$ are the FWHM of the fitted 2D Gaussians. The 1$\sigma$ velocity dispersion $\delta v^{\mbox{\scriptsize{CS}}}$ is measured in a smoothed 20$''$$\times$18$''$ GBT beam area centered at the peak of the fitted 2D Gaussian. The ranges of gas mass in the dense clump are estimated based on the assumption of $\beta$=1-2 (see Section \ref{sub_clump}).
\vspace{0.4cm}}
\end{table*}


\subsection{The Dense Molecular Clumps}
\label{sub_clump}
The brightness temperature of the dust emission at the 0.86 mm wavelength (Appendix \ref{subsub_consistency}) is much lower than the gas temperature (see Herrnstein \& Ho 2005).
The molecular gas mass can therefore be calculated from the 0.86 mm flux based on the optically thin formula
\begin{equation}
M_{\mbox{\scriptsize{H$_{2}$}}} = \frac{2\lambda^{3}Ra\rho D^{2}}{3hcQ(\lambda)J(\lambda,T_{d})}S(\lambda),
\label{eq_mass}
\end{equation}
where $R$ is the gas-to-dust mass ratio, $a$ is the mean grain radius, $\rho$ is the mean grain density, $D$ is the distance of the target, $Q$($\lambda$)$\propto$$\lambda^{-\beta}$ is the grain emissivity, $T_{d}$ is the dust temperature, $S(\lambda)$ is the flux of the dust emission at the given wavelength, $J(\lambda, T_{d})=1/[\mbox{exp}(hc/\lambda k_{B}T_{d})-1]$ (Hildebrand 1983; Lis et al. 1998).
The $c$, $h$, and $k_{B}$ are the light speed, the Planck constant, and the Boltzmann constant, respectively. 
Following Lis at al. (1998), we adopt $R$=100, $a$=0.1 $\mu$m, $\rho$=3 g\,cm$^{-3}$, $Q$($\lambda$=350 $\mu$m)=1$\times$10$^{-4}$.
We adopt $D$=8.33 kpc based on the measurements of Gillessen et al. (2009).

The gas temperature measurements which have a comparable angular resolution with our dust continuum map are not yet available.  
By measuring the rotational temperature of the NH$_{3}$ molecule, Herrnstein \& Ho (2005) suggested that roughly one quarter of the molecular gas comprises a hot ($\sim$200 K) component, and the remaining gas is cool ($\sim$25 K). 
We tentatively assumed an averaged gas temperature T$_{gas}$=70 K in the dense clump.
We also assumed the dust temperature T$_{d}$=T$_{gas}$ (see Chan et al. 1997 for dust temperature in the CND; see also Minh et al.1992, Ao et al. 2012).
The ranges of gas mass in the dense clump estimated based on the assumption of $\beta$=1-2 is given in Table \ref{tab_clump}.
Based on the aforementioned assumptions, without considering the foreground/background subtractions, the detected 0.86 mm flux in the inner 5 pc CND  (68 Jy, see Section \ref{sub_ch0}) corresponds to 0.98-2.3$\times$10$^{4}$ $M_{\odot}$ of gas mass.
Our estimates of the CND mass agree reasonably well with those from earlier observations of millimeter and submillimeter  dust continuum emission (e.g. Mezger et al. 1989; see also Christopher et al. 2005 for relevant debates).

If the identified gas dense clumps are virialized, the expected one-dimensional velocity dispersion $\delta v^{\mbox{\scriptsize{viriral}}}$ is given by
\begin{equation}
\delta v^{\mbox{\scriptsize{viriral}}}= \sqrt{\frac{\alpha MG}{5\delta_{r}}},
\label{eq_virial}
\end{equation}
where $\alpha$ is the geometric factor which equals to unity for a uniform density profile and $5/3$ for an inverse square profile (Williams, de Geus \& Blitz 1994; Walsh et al. 2007), $M$ is the gas mass, $G$ is the gravitational constant, and $\delta_{r}$ is the effective radius of the dense clump. 

The effective radius $\delta_{r}$ can be calculated based on the FWHM of the fitted 2D Gaussian (i.e. $\delta\theta^{\mbox{\scriptsize{maj}}}$ and $\delta\theta^{\mbox{\scriptsize{min}}}$ in Table \ref{tab_clump}), however, is necessary to be corrected for the beam FWHM $\delta\theta^{\mbox{\scriptsize{beam}}}$ because the identified clumps are only marginally resolved. 
For simplicity, we adopt the corrected effective circular angular diameter of the dense clump to estimate their linear diameter (c.f. Williams, de Geus \& Blitz 1994)
\begin{equation}
\delta\theta^{\mbox{\scriptsize{eff}}} =\sqrt{ (\delta\theta^{\mbox{\scriptsize{maj}}}\times\delta\theta^{\mbox{\scriptsize{min}}}) - (2\delta\theta^{\mbox{\scriptsize{beam}}}/2.355)^2}
\label{eq_radius}
\end{equation}
and consider the gas mass enclosed in one FWHM of the fitted 2D Gaussian when calculating the virial one-dimensional velocity dispersion. 
For those identified dense clumps with $\delta\theta^{\mbox{\scriptsize{eff}}}>0$, we compare the results of the calculation with the one-dimensional velocity dispersion measured from the CS 1-0 line and the HCO$^{+}$ 4-3 line (when significantly detected) in Table \ref{tab_clump}.
The mean gas number densities are calculated based on the aforementioned estimates of the clump masses and radii, and the assumptions of spherical geometry and the mean molecular weight of 2.33 (Shull \& Beckwith 1982).
We note that the JCMT-2 region shows double peak profiles in the CS 1-0 spectra. 
However, the broader line component is very faint and therefore is less likely to be associated with the dense gas clumps.

We found that the dense clumps listed in Table \ref{tab_clump} have $\sim$10-10$^{3}$ $M_{\odot}$ of molecular gas, which can be adequate gas reservoirs to form high-mass stars. 
However, the mean gas number density in these dense clumps are generally in between 10$^{5}$-10$^{6}$ cm$^{-3}$, which are marginally unstable against the tidal force (Morris 1993; Liu et al. 2012).
The GBT observations of CS 1-0 suggest that the identified dense clumps are all embedded in very turbulent environment, which have one-dimensional velocity dispersion of 10-30 km\,s$^{-1}$. 
The kinetic energy of the gas may be dissipated on smaller scales. 
The higher angular resolution SMA observations of HCO$^{+}$ 4-3 towards the enclosed Region A, Bn, B, Bs, and JCMT-1n detect the 2-3 times smaller velocity dispersion in the localized dense clumps.
Nevertheless, the observed HCO$^{+}$ 4-3 velocity dispersions are still $\sim$10 times higher than the virial velocity dispersions of the dense clumps.
Our preliminary analyses do not yet find any (self-)gravitationally bound gas structures on $\lesssim$0.5 pc scale. 

Although our HCO$^{+}$ 4-3 image only traces dense clumps in the warmer environments,  HCN 1-0 and HCO$^{+}$ 1-0 spectra from previous 13$''$$\times$4$''$ resolution BIMA array observations also showed broad line profiles in the central 5$'$$\times$5$'$ field (Wright et al. 2001).
Observations of the lower excitation gas tracers HCN 1-0 and HCO$^{+}$ 1-0  trace the velocity dispersion of the dense clumps in the central 5$'$$\times$5$'$ field (Wright et al. 2001).
With 13$''$$\times$4$''$ angular resolution, the BIMA spectra are contaminated by emission from the diffuse and more turbulent ambient gas as traced by CS 1-0, however a deconvolved HCN 1-0 image at $\sim$ 2$''$ resolution (G\"{u}sten et al. 1987b shows only the integrated image) also shows  line widths which are much larger than the virial velocity dispersions of the dense clumps.
Unless these dense clumps are composed of dense, bound cores, or the gas kinetic energy can be dissipated efficiently, they may be dispersed in one dynamical timescale (i.e. $\delta_{r}/\delta v$$\sim$5.3$\cdot$10$^{3}$-2.6$\cdot$10$^{4}$ years for the HCO$^{+}$ 4-3 emission clumps in Table 1).
Higher angular and velocity resolution spectral line observations are required to see whether (self-)gravitationally bound gas structures exist on a smaller scale.

Alternatively, these dense clumps embedded in the extremely turbulent molecular clouds may be confined by the high external pressure.
For example, the previous X-ray studies (e.g. Koyama et al. 1996, Muno et al. 2004) have demonstrated that a high-pressure medium with sound speed $>$1000 km\,s$^{-1}$ pervades the Galactic center region.
If small scale virialized gas structures do exist inside the pressurized dense clumps, then hydrostatic cores and stars may still form.
Assuming approximate hydrostatic equilibrium of the embedded star-forming cores with masses $M_{core}$, the radius $r_{s}$, the mean gas number density $\bar{n_{H}}$, and the r.m.s velocity dispersion $v_{rms}$ of the cores, as well as the final stellar mass $m_{*f}$=$\epsilon_{core}$$M_{core}$, can be estimated based on the following formulae (details see McKee \& Tan 2002):
\begin{equation}
r_{s} = 0.074(m_{*f}/30 M_{\odot})^{1/2}\Sigma^{-1/2}\,\,\,\mbox{pc},
\end{equation}
\begin{equation}
\bar{n_{H}} = 1.0\times10^{6}(m_{*f}/30 M_{\odot})^{-1/2}\Sigma^{3/2}\,\,\,\mbox{cm$^{-3}$},
\end{equation}
\begin{equation}
v_{rms} = 1.65(m_{*f}/30 M_{\odot})^{1/4}\Sigma^{1/4}\,\,\,\mbox{km\,s$^{-1}$},
\end{equation}
where $\Sigma$ is the mean clump surface mass density in units of g\,cm$^{-2}$, and $\epsilon_{core}$ (assumed to be 0.5 here) is the fraction of the core mass which is eventually accreted onto the central star.
From Figure \ref{fig_ch0ffsub}, the majority of the densest clumps are embedded in the region above the 12$\sigma$ contour level (i.e. 288 mJy\,beam$^{-1}$), which corresponds to the $\Sigma$ of 0.31-0.76 g\,cm$^{-2}$. 
Based on the values in Table \ref{tab_clump}, we adopt a fiducial value $\bar{n_{H}}$$\sim$10$^{6}$ cm$^{-3}$ for an example.
We estimate the core radius $r_{s}$$\sim$0.023-0.056 pc (0.58$''$-1.4$''$), the r.m.s. velocity dispersion 0.51-1.3 km\,s$^{-1}$, and the final stellar mass $m_{*f}$$\sim$0.89-13 $M_{\odot}$.
The former two quantities can be examined with $<$0.5$''$ resolution Atacama Large Millimeter/submillimeter Array (ALMA) observations in the near future.
The corresponding core mass $M_{core}$=2$\times$$m_{*f}$ in our estimates is several times smaller than the masses of the parent gas clumps listed in Table \ref{tab_clump} so may be reasonable.

As discussed in the previous section, the dense clumps embedded in some regions (e.g. A, B, D, E, F, G) are associated with the 22 GHz water maser and the 36.2 GHz and 44.1 GHz Class I CH$_{3}$OH masers, which are often seen in the early phase of high-mass star-formation. 
In addition, the earlier VLA observations of the 2 cm and the 6 cm continuum emission have found ultracompact H\textsc{ii} regions embedded with several OB stars in the east of the 50 km\,s$^{-1}$ cloud (e.g. Ho et al. 1985).
The observed high HCO$^{+}$ 4-3 velocity dispersion may be interpreted by (proto)stellar feedback. 
However, Sjouwerman et al. (2010) also suggested that the 36.2 GHz Class I CH$_{3}$OH masers can be excited at the post shock regions created by cloud-cloud collisions.
High angular resolution molecular line observations and studies of the maser proper motion may distinguish these two cases, although they are not mutually exclusive. 
In fact, it has been argued that the predominant mode of star formation is via external compression of molecular clouds by cloud collisions and supernova (Morris 1993).
The high-mass star formation can also be induced by AGN activities (Silk et al. 2012).
In these cases, the physical properties of the molecular gas reservoir feeding the high-mass star formation may be very different from the typical OB star-forming cores.

\subsection{The Non-Uniform CND}
\label{sub_cnd}
\begin{figure}[h]
\vspace{-0.5cm}
\includegraphics[width=9cm]{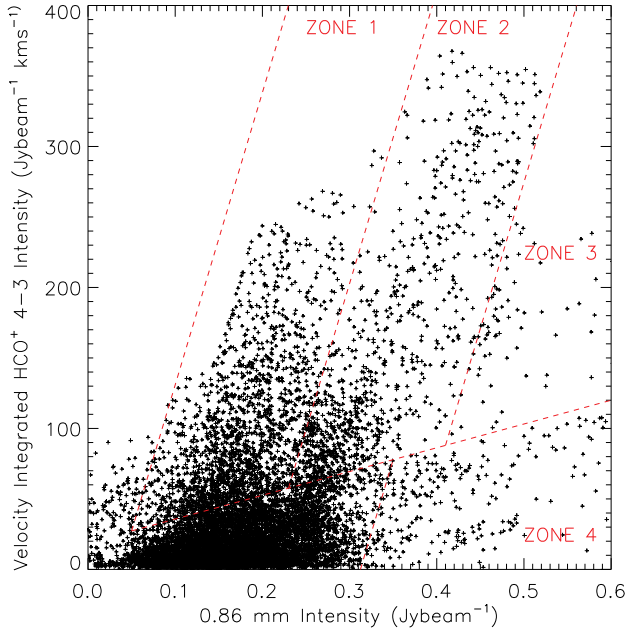}
\vspace{-0.6cm}
\caption{\footnotesize{
The velocity integrated HCO$^{+}$ 4-3 intensity and the 0.86 mm intensity of pixels within the 60$''$ radius around the Sgr A*.
We color code the pixels in the four selected zones in Figure \ref{fig_zone}.
We note that our HCO$^{+}$ 4-3 image is subjected to missing flux, however, should not significantly bias the analysis of the emission from the localized clumps. 
}}
\vspace{0.5cm}
\label{fig_diagram_hcop}
\end{figure}

Previous interferometric spectral line observations have shown abundant localized structures in the central 2-4 pc CND (see Christopher et al. 2005; Montero-Casta{\~n}o et al. 2009; Mart{\'{\i}}n et al. 2012, and references therein).
The distributions of the spectral line emission, however, are sensitive to the local temperature,
volume density,  chemical abundances, and other radiation transfer effects such as the foreground absorption or self-absorption. 
The optically thin 0.86 mm dust thermal continuum emission is a more robust tracer of the gas mass (e.g. Hildebrand 1983). 
Without being subjected to missing flux, our JCMT+SMA 0.86 mm dust continuum image (Figure \ref{fig_ch0ffsub}) successfully reproduces the clumpy CND structures seen in the spectral line observations (e.g. Figure \ref{fig_hcop}).
It appears that some previously known clumps in the CND are not merely due to excitation or radiation transfer effects. 
For example, the 0.86 mm emission clump A and Bn (Table \ref{tab_clump}; Figure \ref{fig_regions}) coincide with the HCN 4-3 emission clump CC and BB reported by Montero-Casta{\~n}o et al.(2009).
As can be expected, we found that around the CND, the velocity integrated HCO$^{+}$ 4-3 intensity is correlated with the intensity of the 0.86 mm continuum emission (Figure \ref{fig_diagram_hcop}).

However, in Figure \ref{fig_diagram_hcop}, we see that the correlation between the HCO$^{+}$ 4-3 emission and the 0.86 mm emission is dominated by at least two distinct populations which can be discerned with our current angular resolution and sensitivity. 
We selected the high HCO$^{+}$ 4-3 intensity parts of the two dominant populations as Zone 1 and Zone 2 in Figure \ref{fig_diagram_hcop}, and additionally one high HCO$^{+}$ 4-3 intensity and high 0.86 mm intensity population as Zone 3, and one high 0.86 mm intensity population as Zone 4.
We argue that these high intensity pixels are more likely to be associated with compact bright clumps, for which the HCO$^{+}$ 4-3 intensity is less biased by missing flux. 

\begin{figure}[h]
\vspace{-1.5cm}
\begin{tabular}{c}
\\
\end{tabular}
\includegraphics[width=9cm]{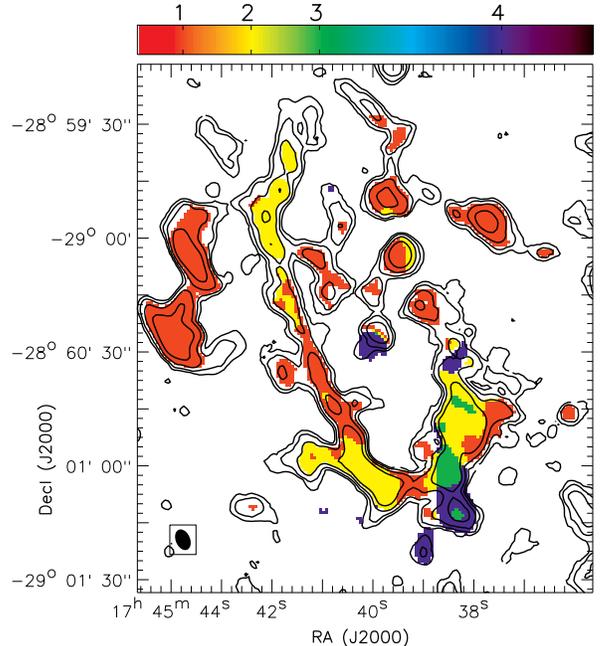}
\vspace{-0.6cm}
\caption{\footnotesize{
The color coded pixels which reside at Zone 1 (red), 2 (yellow), 3 (green), and 4 (purple) in Figure \ref{fig_diagram_hcop} respectively.
The color image is overlaid with the velocity integrated HCO$^{+}$ 4-3 image (contours; $\theta_{\mbox{\scriptsize{maj}}}\times\theta_{\mbox{\scriptsize{min}}}$=5$''$.8$\times$4$''$.1; see Figure \ref{fig_hcop}).
Contours are 3$\sigma$$\times$[1, 2, 4, 8] ($\sigma$=15.5 mJy\,beam$^{-1}$km\,s$^{-1}$).
}}
\vspace{0.0cm}
\label{fig_zone}
\end{figure}

The spatial distributions of the pixels in these four zones are not random.
For example, the pixels in Zone 1 seems to trace the two arc-shaped clumps west of the CND, the eastern edge of the CND, and the clumps located northwest of the Sgr A*.
The pixels in Zone 2 are associated with the Northeast Lobe, the Southern Extension, and northern part of the Southwest Lobe.
The southern part of the Southwest Lobe is associated with the pixels in Zone 4.
Comparing Figure \ref{fig_zone} with the ratio of the HCN 4-3 to HCN 1-0 integrated intensity reported by Montero-Casta{\~n}o et al. (2009), we deduce that the very dense clump in the southern part of the Southwest Lobe has a low gas excitation temperature.
This provide hints for the the cooler exterior gas clumps raining down on the warmer CND.
For a dynamically evolving CND which is connected with several exterior gas streamers (see Liu et al. 2012), it is not surprising that the excitation conditions, the chemically abundances, or the dust properties are not yet homogenized.

\section{Summary}
\label{chap_summary}
We present a wide-field SMA mosaic toward the Galactic center.
We also observed the CS/C$^{34}$S/$^{13}$CS 1-0 and the SiO 1-0 lines using the GBT.
The optically thin 0.86 mm dust thermal continuum image with $\sim$5$''$ angular resolution confirms the 2-10 pc scale gas streamers and the detailed structures of the CND, which were seen in previous molecular line observations.  
We marginally resolve more than 22 massive (10$^{1}$-10$^{3}$ $M_{\odot}$) gas clumps, which are embedded in very turbulent molecular gas clouds ($\delta v\sim$10-30 km\,s$^{-1}$). 
Examination of the brightness ratio of the HCO$^{+}$ 4-3 line and the 0.86 mm continuum emission shows that the non-homogenized central 2-4 pc CND has dense clumps with distinct excitation conditions or chemical abundances.
While the distributions of the 1720 MHz and 1612 MHz OH maser clusters do not show obvious correlations with the dust emission, the 22 GHz water masers and especially the 36.2 and 44.1 GHz Class I CH$_{3}$OH masers are seen preferentially near the dense clumps.
Even the most significant dense clumps in our selected samples are marginally unstable against the tidal force, and 
 we do not find any self-gravitationally bound gas clump associated with the Class I CH$_{3}$OH masers or the 22 GHz H$_{2}$O masers. 
How the OB stars form in such environment remains puzzling. 
The mechanisms to form the dense clumps and the high-mass stars in the Galactic center might be very different from the mechanisms in  typical giant molecular clouds.
Our simple estimates suggest that if the detected dense clumps are confined by the high external pressure, the presumably existing embedded virialized gas cores can form high-mass stars.
Deep JVLA observations to search for the signature of the stellar photoionization, or ALMA observations of the high excitation hot core tracers to look for the gravitationally accelerated rotation, may diagnose the OB star formation in the dense clumps. 

Our GBT data suggest  mildly enhanced SiO/C$^{34}$S line ratio towards the Southern Arc and the 20 km\,s$^{-1}$ cloud, relative to the SiO/C$^{34}$S line ratios in the other gas streamers in the observed field. 
Our observations do not yet resolve the recognizable correlation between the SiO/C$^{34}$S line ratio and the intensity of the submillimeter dust emission.
Higher angular resolution and more sensitive observations are required to examine whether the formation of the dense gas structures are predominantly induced by shocks.

\acknowledgments 
The GBT and SMA data are from projects GBT11BÐ050, SMA2011AÐS085 and SMA2011BÐS040, which are parts of the integrated state-of-art imaging project KISS: \underline{K}inematic Processes of the Extremely Turbulent \underline{IS}M around the \underline{S}upermassive Black Holes. 
We acknowledge financial support from ASIAA. 
S.S.K. was supported by Mid-career Research Program (No. 2011-0016898) through the National Research Foundation (NRF) grant funded by the Ministry of Education, Science and Technology (MEST) of Korea.
We thank Dr. Eric Feigelson for the very useful suggestions.
We thank Zhao Jun-Hui very much for his efforts in optimizing and upgrading MIRIAD for SMA, which made this project possible.
We thank Toney Minter, Glen Langston, and David T. Frayer for assisting the GBT observations. 
We thank Glen Petitpas and Nimesh Patel for supporting the SMA observations.  
This research used the facilities of the Canadian Astronomy Data Centre operated by the National Research Council of Canada with the support of the Canadian Space Agency. 
{\it Facilities:} \facility{SMA, GBT}



\appendix


\section{A. Short Spacing Information}
\label{subsub_zero}
\begin{figure}[h]
\hspace{-1cm}
\begin{tabular}{c}
\vspace{-2cm} \\
\includegraphics[width=10.5cm]{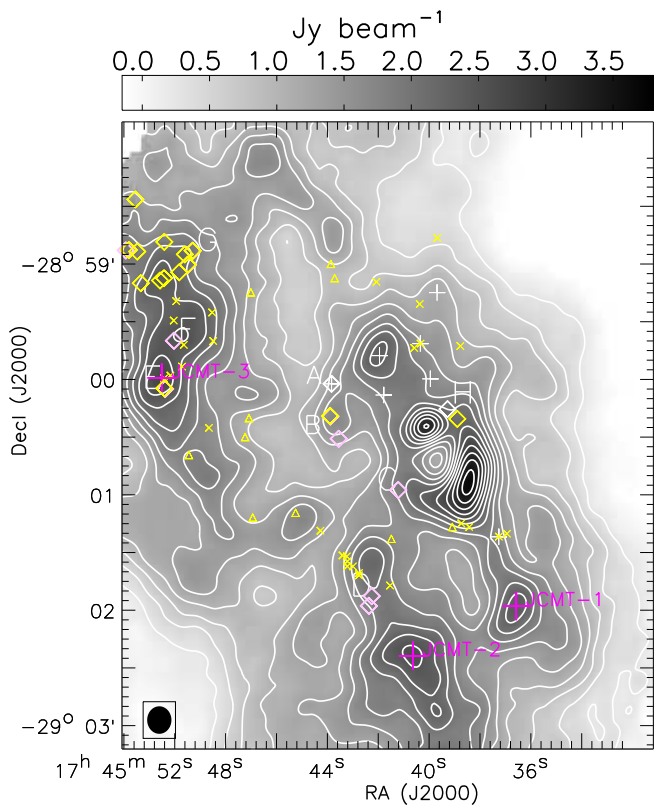} \\
\end{tabular}
\vspace{-0.4cm}
\caption{\footnotesize{
The JCMT SCUBA 0.86 mm continuum image ($\theta_{\mbox{\scriptsize{maj}}}\times\theta_{\mbox{\scriptsize{min}}}$=14$''$.3$\times$14$''$.3).
Contour spacings are 5$\sigma$ starting at 5$\sigma$ ($\sigma$=50 mJy\,beam$^{-1}$).
For our convenience in discussion, we mark the peaks above the 45$\sigma$ significance level besides the Northeast Lobe and Southwest Lobe as JCMT-1,2,3.
The symbols are described in Figure \ref{fig_ch0ffsub}.
}}
\label{fig_jcmt}
\end{figure}

Short spacing information, below $\sim$7.0 $k\lambda$, was obtained from single dish telescope observations. 
We retrieved the archival public processed JCMT\footnote{The James Clerk Maxwell Telescope is operated by the Joint Astronomy Centre on behalf of the Science and Technology Facilities Council of the United Kingdom, the Netherlands Organisation for Scientific Research, and the National Research Council of Canada.} SCUBA (ProjectID: M01BU26, observed on 2001 August 05) 0.86 mm continuum image (Figure \ref{fig_jcmt}; rms$\sim$50 mJy\,beam$^{-1}$).
We used the MIRIAD task \texttt{DEMOS} to generate SMA primary beam weighted models for each SMA pointing.
For each SMA field, we then resampled the primary beam weighted models in the uv domain with 185 Gaussian randomly distributed visibilities using the MIRIAD tasks \texttt{UVRANDOM} and \texttt{UVMODEL}.
We manually assigned a system temperature of 350 K to the visibility model to adjust the weighting.

The typical pointing accuracy for JCMT SCUBA is about 3$''$ and the tracking accuracy  $\sim$1.5$''$ (Di Francesco et al. 2008). 
From their JCMT SCUBA legacy survey, Di Francesco et al. (2008) also mentioned that larger pointing offsets (e.g. $\sim$6$''$) occurred occasionally.
When combining with the interferometric observations, pointing offsets during the single dish observations can cause defects especially near  bright compact sources.
We used a limited uv  range of 0-3.8 k$\lambda$ for the single dish uv model to suppress  the potential effect of the single dish pointing offsets, as well as the effect of the single dish primary beam and deconvolution errors.
Although the effect of the single dish primary beam is not fully eliminated, it is smaller than the $\sim$20\% absolute flux error in typical SMA observations  (Appendix \ref{subsub_consistency}).
Limiting  the uv  range implies that the single dish primary beam causes defects primarily for structures  $>$33$''$, and therefore should not confuse the identification of local gas clumps in the streamers.

We jointly deconvolved and imaged the concatenated single dish uv model and the SMA data using the MIRIAD software package (using tasks: \texttt{INVERT}, \texttt{MOSSDI}, \texttt{RESTOR}). 
The gap (3.8-7 k$\lambda$) in the uv sampling may yield uncertainties in reconstructing the 14$''$-33$''$ angular scale structures.
However, structures with these angular scales can be directly inspected from the single dish image.
We generated the high angular resolution continuum image with the parameters \textit{robust=0 fwhm=4,4} in \texttt{INVERT}, which yield a synthesized beamwidth $\theta_{\mbox{\scriptsize{maj}}}\times\theta_{\mbox{\scriptsize{min}}}$=5$''$.1$\times$4$''$.2, and BPA. $\sim$28$^{\circ}$.
To compare with the earlier observations, we also generated one lower angular resolution ($\theta_{\mbox{\scriptsize{maj}}}\times\theta_{\mbox{\scriptsize{min}}}$=7$''$.4$\times$6$''$.2, BPA. $\sim$44$^{\circ}$) continuum image with the weighting parameters \textit{robust=0 fwhm=8,8} to enhance the sensitivity to the extended emission.

\begin{figure}
\hspace{-1cm}
\begin{tabular}{cc}
\vspace{-2cm} \\
\includegraphics[width=10.5cm]{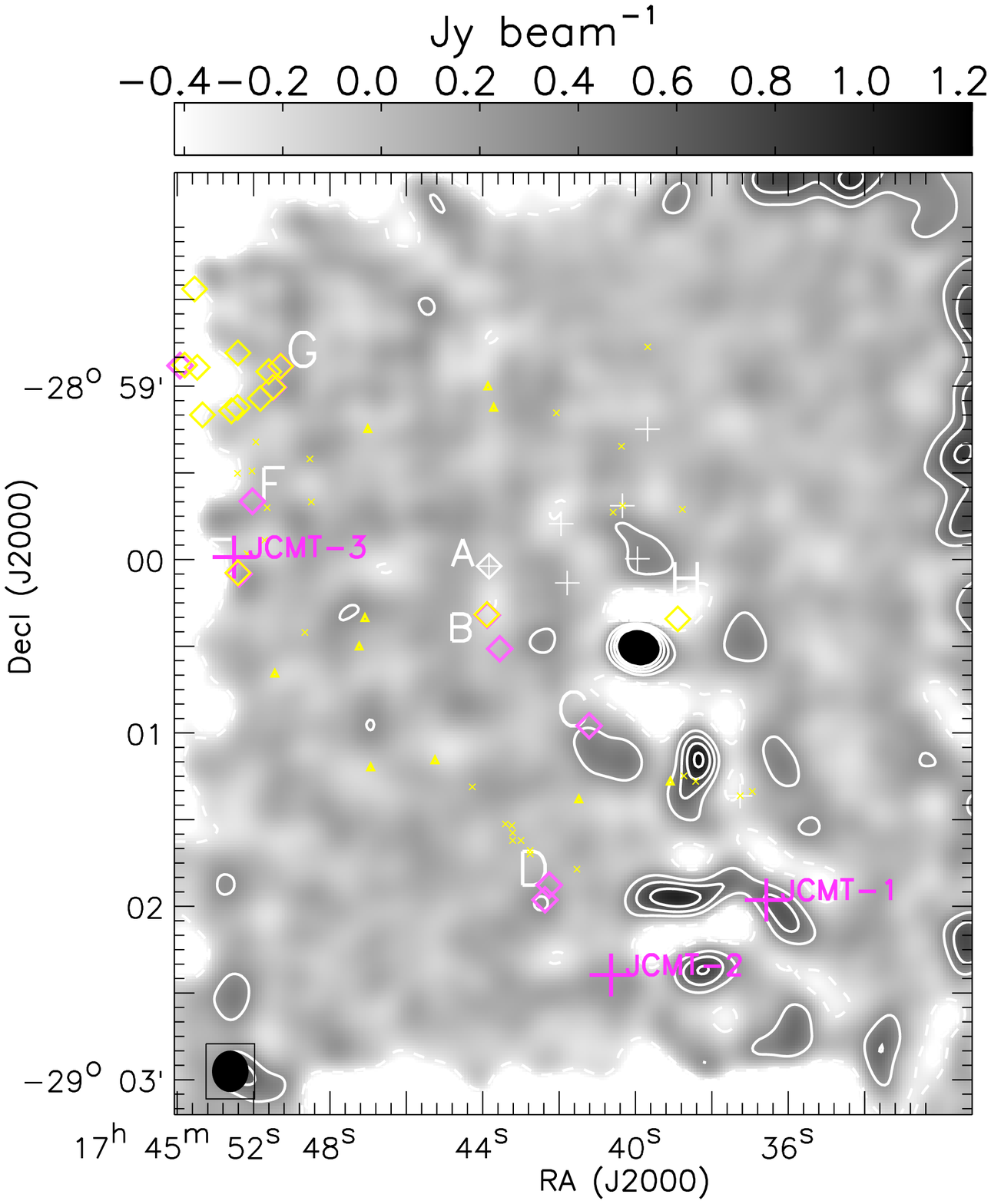}  & \hspace{-2cm}\includegraphics[width=10.5cm]{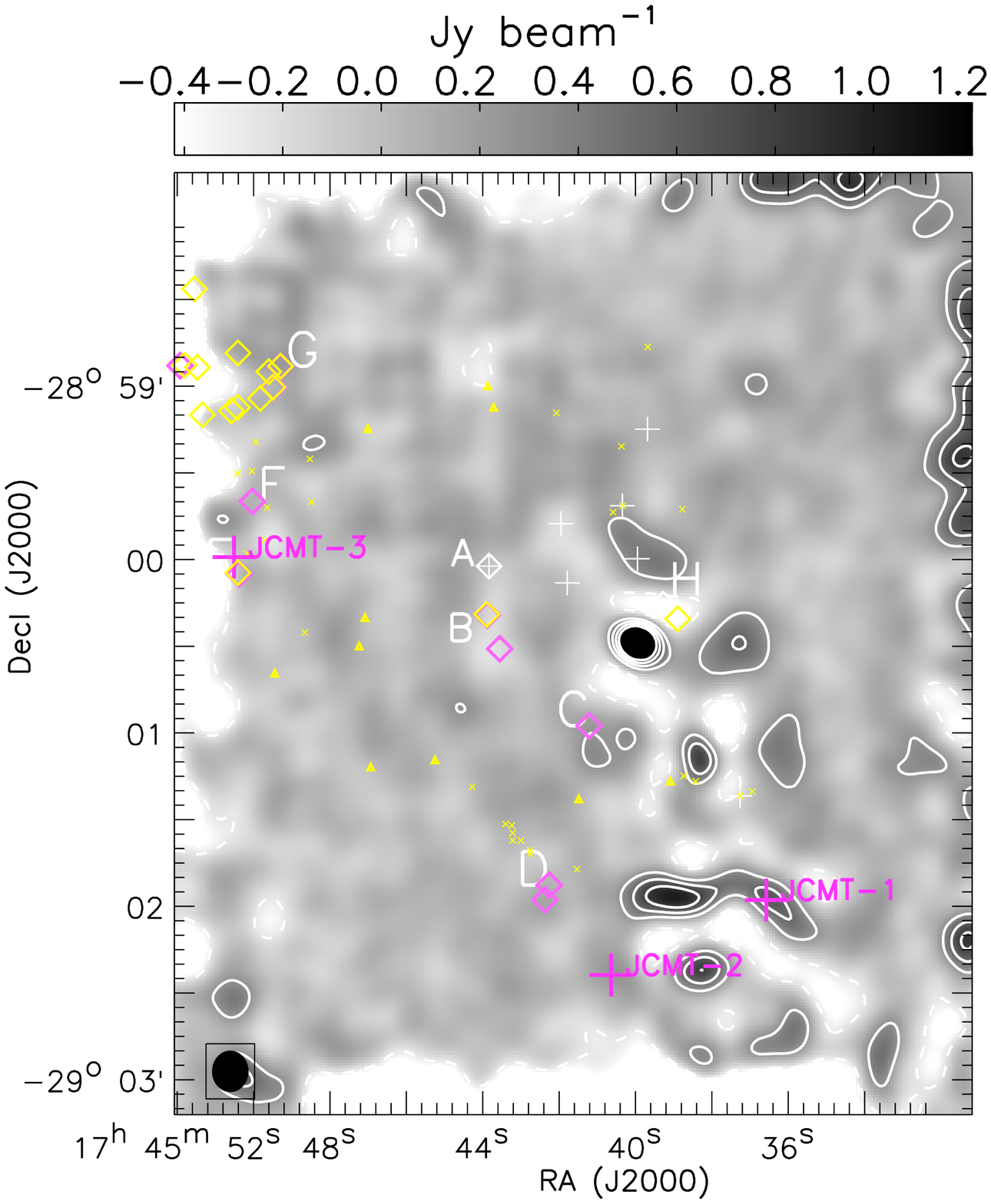} \\
\end{tabular}
\vspace{-0.4cm}
\caption{\footnotesize{
Left: the residual image constructed by subtracting the JCMT SCUBA image from the smoothed high angular resolution SMA+JCMT 0.86 mm continuum image (see Appendix \ref{subsub_consistency}).
The JCMT beam is shown in the lower left.
Contours are 3$\sigma$$\times$[-2, -1, 1, 2, 3, 4, 5] ($\sigma$=89 mJy\,beam$^{-1}$).
We note that the averaged 3$\sigma$ flux density 267 mJy\,beam$^{-1}$ in this image corresponds to the averaged flux of $\sim$28 mJy in each 5$''$.1$\times$4$''$.2 SMA synthesized beam.
The symbols are described in Figure \ref{fig_ch0ffsub}.
Right: similar to the left panel. However, the original JCMT SCUBA image was shifted toward the south by 4$''$ before being subtracted from the SMA+JCMT image.
}}
\label{fig_consistency}
\end{figure}


\section{B. Consistency Check and Noise Statistics}
\label{subsub_consistency}
\begin{figure*}
\vspace{-1.5cm}
\begin{tabular}{c}
\\
\end{tabular}

\hspace{-1cm}
\begin{tabular}{ p{9.2cm} p{9.2cm} }
 \includegraphics[width=10.5cm]{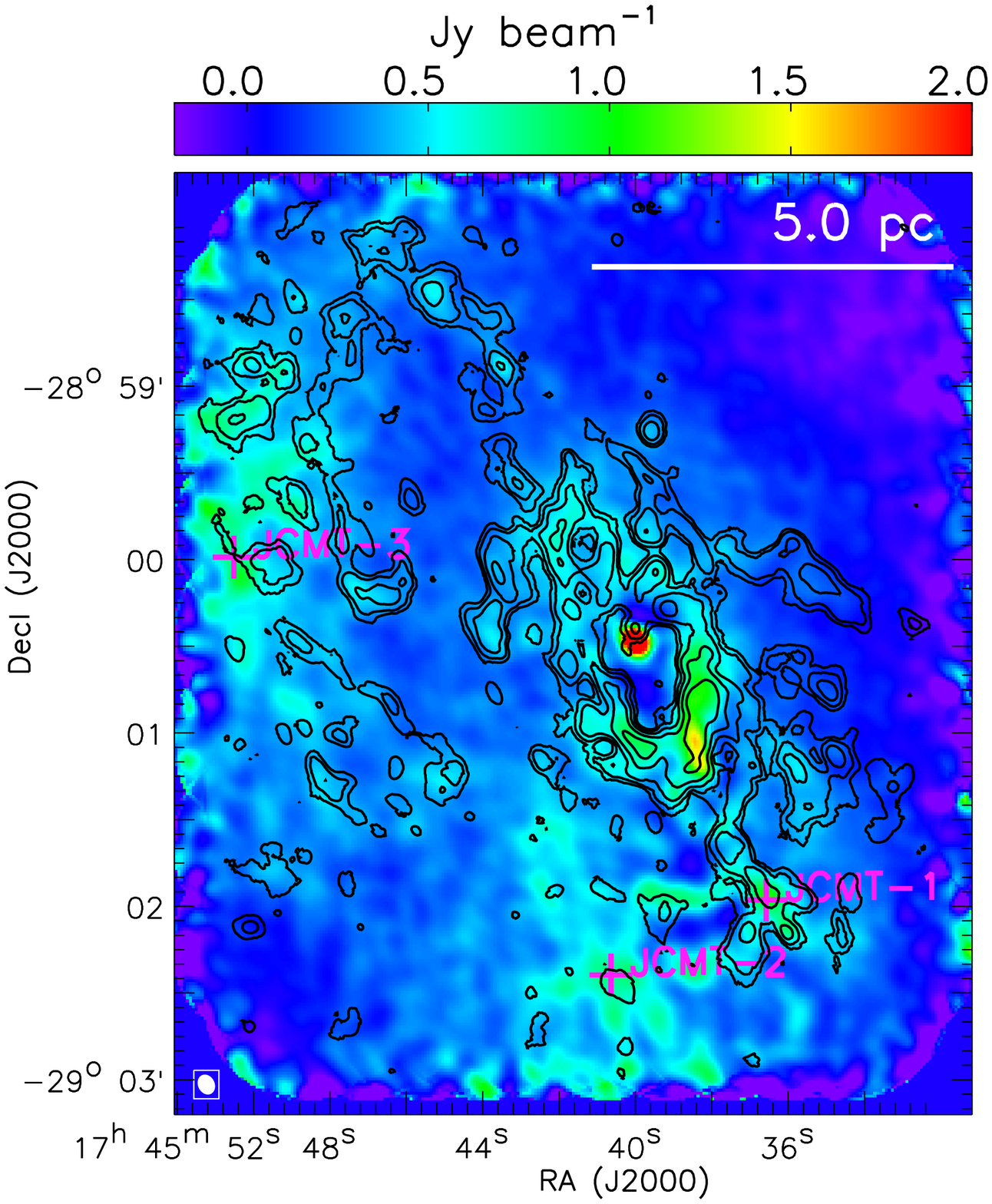} & \includegraphics[width=10.5cm]{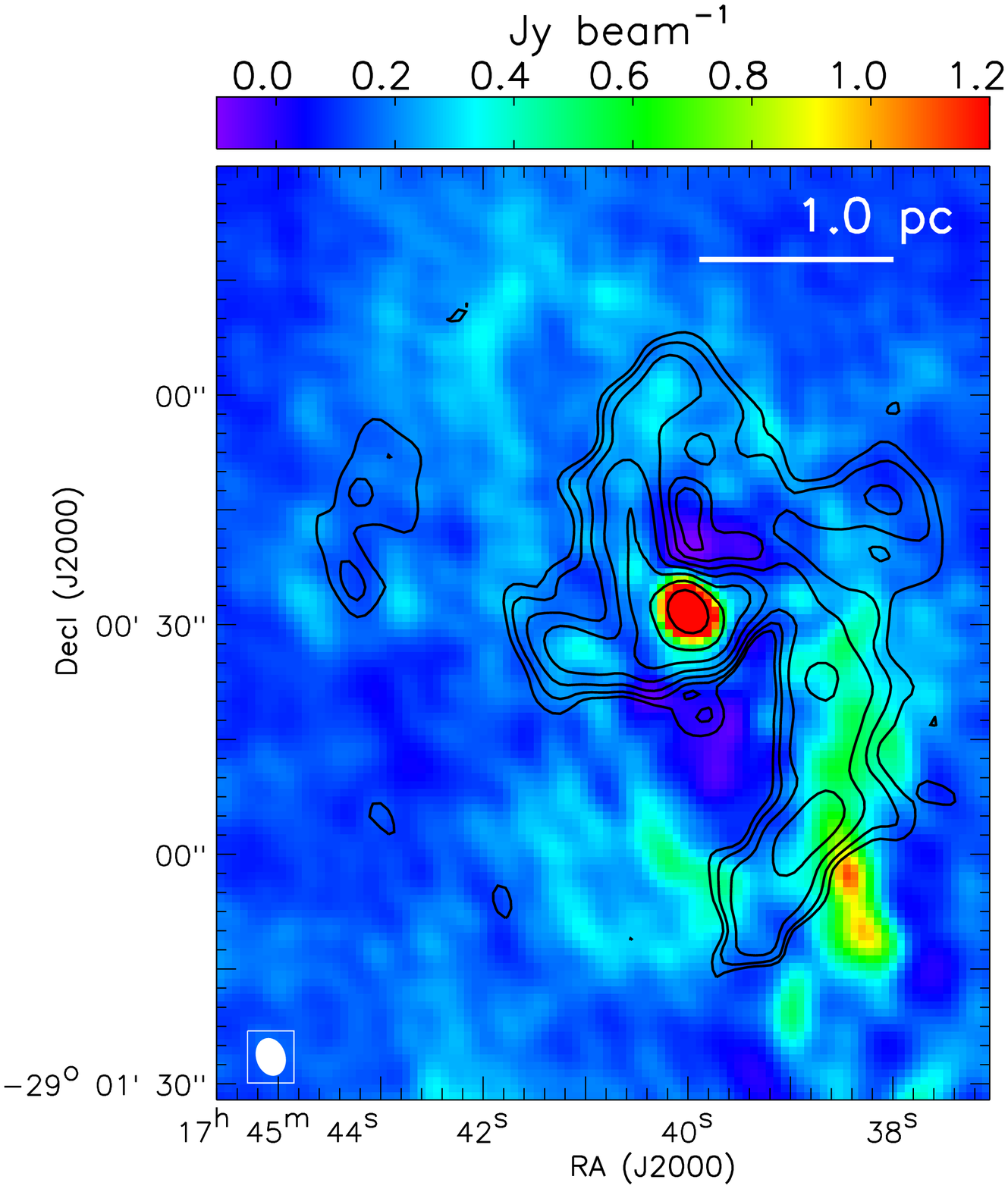}\\
\end{tabular}
\caption{\footnotesize{
The combined SMA+JCMT 0.86 mm continuum image (color).
\textit{Left: } Color image shows the tappered SMA+JCMT 0.86 mm continuum image ($\theta_{\mbox{\scriptsize{maj}}}\times\theta_{\mbox{\scriptsize{min}}}$=7$''$.4$\times$6$''$.2). The synthesized beam of the SMA+JCMT image is shown in the lower left. Contours show the velocity integrated SMA HCN 4-3 image taken from Liu et al. (2012)  ($\theta_{\mbox{\scriptsize{maj}}}\times\theta_{\mbox{\scriptsize{min}}}$=5$''$.9$\times$4$''$.4).  Contours  are 75 Jy\,beam$^{-1}$km\,s$^{-1}$$\times$[1, 2, 4, 8, 16]. By comparing with the GBT CS 1-0 spectra, Liu et al. (2012) suggested that this HCN 4-3 image is subjected to the significant missing flux issue around the 50 km\,$^{-1}$ cloud, the Molecular Ridge, the 20 km\,s$^{-1}$, and the Southern Arc. 
\textit{Right: } Color image shows the high angular resolution SMA+JCMT 0.86 mm continuum image ($\theta_{\mbox{\scriptsize{maj}}}\times\theta_{\mbox{\scriptsize{min}}}$=5$''$.1$\times$4$''$.2). Contours show the free-free continuum emission model (see Appendix \ref{subsub_freefree}), which is smoothed to the same angular resolution with the SMA+JCMT image. Contours are 10 mJy\,beam$^{-1}$$\times$[1, 2, 4, 8, 16, 32, 64].
}}
\label{fig_ch0}
\end{figure*}

We smoothed the high angular resolution SMA+JCMT image to the angular resolution of JCMT SCUBA, and then subtracted the JCMT SCUBA image from the smoothed SMA+JCMT image. 
The residual image after the subtraction is shown in Figure \ref{fig_consistency}.
The lower angular resolution SMA+JCMT image, and the previously published SMA HCN 4-3 image (Liu et al. 2012) are presented in Figure \ref{fig_ch0}.
On average the contour intervals in Figure \ref{fig_consistency} (i.e. 267 mJy\,beam$^{-1}$) correspond to the 1$\sigma$ (24 mJy\,beam$^{-1}$) detection level of the higher angular resolution ($\theta_{\mbox{\scriptsize{maj}}}\times\theta_{\mbox{\scriptsize{min}}}$=5$''$.1$\times$4$''$.2) SMA+JCMT 0.86 mm continuum image (see below).

Our overall impression by comparing Figure \ref{fig_consistency} and \ref{fig_ch0} is that the limited uv sampling rate of our SMA observations cause imaging defects near the bright compact sources. 
This is most obviously seen around the Sgr A*, the Southwest Lobe, and the bright clump in the Southern Arc (coincide with JCMT-1).
However, we found that manually shifting the JCMT image towards the south by 4$''$ can suppress the residual (e.g. Figure \ref{fig_consistency}) around the Sgr A*.
In our zero spacing model (see Section \ref{subsub_zero}), we did not manually correct for this 4$''$ offset for the sake of objectiveness. 
We do not think this potential JCMT 4$''$ pointing offset can cause very obvious imaging defects since only the $\ge$33$''$ scale structures are taken for modeling the zero spacing.
However, it can lead to the mismatch between the combined SMA+JCMT image and the JCMT image, which will result in some residual in Figure \ref{fig_consistency}.

Beyond the aforementioned regions, the extended structures appear to be reasonably reconstructed in the SMA+JCMT image, and the residual image has an rms noise level of $\sim$89 mJy\,beam$^{-1}$.
The Northern/Southern Ridges, Northeast/Southwest Lobes, and the W-2,3,4 Streamers are detected in the lower angular resolution SMA+JCMT image (Figure \ref{fig_ch0} left).
The extended residual features at the edge of Figure \ref{fig_consistency} are caused by the sidelobe features in the JCMT SCUBA image (e.g. the negative brightness sidelobe at northwest of Figure \ref{fig_jcmt} causes the positive excess in Figure \ref{fig_consistency}).
These single dish sidelobes are extended ($\ge$15$''$) and should not confuse the imaging of localized gas structures.
However, locally they can bias the absolute brightness in the higher angular resolution SMA+JCMT image (mostly only 2$\sigma$).
The $>$33$''$ scale defects caused by the single dish primary beam effect discussed in the previous section appear lower than the 1$\sigma$ detection level of the SMA+JCMT image.

The theoretical rms noise level of the SMA+JCMT image is $\sim$14 mJy\,beam$^{-1}$.
Because of the absence of emission-free areas in our map, it is difficult to directly measure the rms noise level we actually achieved. 
Since only the $\ge$33$''$ structures in the JCMT image, which have high signal to noise ratios, are combined with the SMA data.
We assumed the noise in the SMA+JCMT image only has a weak dependence on the noise in the JCMT image.
We therefore can measure the rms noise level of the high angular resolution SMA+JCMT image by
\begin{equation}
\frac{ \sqrt{(\mbox{\scriptsize{rms of the residual image}})^2 - (\mbox{\scriptsize{rms of the JCMT image}})^2} }{ \sqrt{\mbox{\scriptsize{(JCMT beam area)}} / \mbox{\scriptsize{(SMA+JCMT synthesized beam area)}} } }.
\end{equation}
This yields an rms noise level of 24 mJy\,beam$^{-1}$ (0.011 K) in the high angular resolution SMA+JCMT image, which is $\sim$1.7 times higher than the theoretical noise level.
Generally, the SMA+JCMT image is still dynamic-range limited, especially near the bright compact objects. 
This effect hampers the systematical search and statistical study of  the clumpy structures.

\section{C. Free-free Model Subtraction}
\label{subsub_freefree}
The emission from the ionized mini-spiral arms can be recognized from the high angular resolution SMA+JCMT image (Figure \ref{fig_ch0}).
We retrieved the archival VLA Q band observations taken on 2003 February 14, March 17, April 16, and 2004 March 14, April 21, May 16 to generate a model of 0.86 mm free-free emission.  
These VLA observations were taken with dual 50 MHz IFs centered at 43.314 GHz and 43.364 GHz, with full polarization.
The sampling range of these VLA data, is 3.7-490 k$\lambda$.
However, we assume that only the localized bright components will significantly contribute to the 0.86 mm free-free continuum emission. 
The basic calibrations and self-calibrations were performed using the Astronomical Image Processing System (AIPS) software package of NRAO. 
We scaled the VLA image such that the central point source Sgr A* has the flux of 1 Jy (Bower \& Backer 1998; Yusef-Zadeh et al. 2011), and then smoothed the VLA image to the angular resolution of the SMA+JCMT image.
The smoothed and rescaled VLA image, as the zeroth order model of the 0.86 mm free-free continuum emission, is presented in the right panel of Figure \ref{fig_ch0}.
The brightness of the 3 mm and the 1.3 mm emission is comparable, which can be checked in Kunneriath et al. (2012).
We then subtracted the free-free emission model from the SMA+JCMT image (Figure \ref{fig_ch0ffsub}).
As can be seen from Figure \ref{fig_ch0} and \ref{fig_ch0ffsub}, this free-free model subtraction can only manifestly change the geometry in the Northern and the Eastern ionized mini-spiral arm regions and around Sgr A* (see Zhao et al. 2009 and references therein for the ionized mini-spiral arm).
We note that the scaling of the VLA image in this process cannot take into the consideration of the spatial variation of the spectral index (Kunneriath et al. 2012). 
We have checked that if we scaled the VLA image to be $\ge$20\% brighter, this free-free model subtraction will induce noticeable over-subtracted features in the Northern and the Eastern ionized mini-spiral arms.
Except for the location of Sgr A*, in the central 1$'$ area, the free-free model subtracted SMA+JCMT image provides a lower limit for the 0.86 mm dust thermal emission.

\end{document}